\documentclass[1p,review,numbers,sort&compress]{elsarticle}
\usepackage{amsmath,amssymb,graphicx,color,here,subfig,natbib,hyperref}
\usepackage{upgreek}

\makeatletter
\def\ps@pprintTitle{%
	\let\@oddhead\@empty
	\let\@evenhead\@empty
	\def\@oddfoot{\reset@font\hfil\thepage\hfil}
	\let\@evenfoot\@oddfoot}
\makeatother

\providecommand{\bibinfo}[2]{#2}
\providecommand{\doi}[1]{doi:\,#1}

\begin{document}
	
	\begin{frontmatter}
		
		\title{Why fluid relaxation modes matter: Nonlinear frequency response, dynamical integrity and dynamic pull-in of electrostatic microbeams with squeeze-film damping}
		
		\author[els]{Fehmi NAJAR \corref{cor1}}
		\cortext[cor1]{Corresponding author, e-mail: f.najar@psau.edu.sa}
		\address[els]{Department of Mechanical Engineering, College of Engineering, \\
			Prince Sattam bin Abdulaziz University, Al-Kharj 16273, Saudi Arabia.}

		\begin{abstract}
			The squeeze film between an electrostatically actuated microbeam can both induce damping and add stiffness to the structure. In addition, the dynamics of the film can interact with the nonlinear response of the microbeam. This paper extends a reduced-order model built on the complex modes of the coupled fluid--structure problem to the computation of periodic responses and their stability, for clamped--clamped microbeams. Periodic orbits are continued in the frequency domain and their stability is determined from their Floquet multipliers. The robustness of the response is assessed through basins of attraction, integrity measures and dynamic pull-in maps in the plane of the excitation parameters. The model is validated against the measured and computed frequency response of a microswitch. Closed-form expressions are derived for the fluid relaxation modes, together with a correction for strongly coupled films, and define a parameter that characterizes the regime of the film. The approach is applied to different microbeams including cases where fluid modes relax on the timescale of the oscillation. For all simulated devices, a single fluid mode is necessary and sufficient, whereas adding coupled modes cannot replace it, whether the fluid modes respond quasi-statically or carry their own dynamics. Removing the fluid mode displaces both bifurcations that bound the pull-in band, changes the size of the basins of attraction and the position of the attractor within them, misses an abrupt erosion of the safe basin under increasing excitation, and shifts the whole region of dynamic pull-in in the excitation plane. The reduced model, with three real unknowns, makes these global analyses affordable for a fully coupled fluid--structure system.
		\end{abstract}
		
		
	\end{frontmatter}

% ---------------------------------------------------------------------------
\section{Introduction} \label{sec:intro}

Electrostatically actuated microbeams are the building blocks of many microelectromechanical systems, such as resonant sensors, filters and radio-frequency switches. Their response to a combined DC and AC voltage is strongly nonlinear. The quadratic dependence of the electrostatic force on the voltage and its inverse dependence on the gap, combined with the mid-plane stretching of clamped microbeams, lead to complex dynamics. Softening and locally hardening frequency responses, superharmonic and subharmonic resonances, period-doubling bifurcations and dynamic pull-in, the collapse of the microbeam onto the electrode at voltages well below the static pull-in voltage are examples of these complexities \citep{nayfeh2005reduced,nayfeh2007dynamic,najar2010nonlinear,alsaleem2009nonlinear}. In particular, dynamic pull-in is a failure mode for resonant sensors and filters, but it is also the operating principle of low-voltage microswitches actuated by a combined DC and AC loading, as proposed by \citet{nayfeh2007dynamic}, which switch at lower voltages and with waveforms better suited to integrated circuits \citep{sassi2018square}. In both cases, predicting the frequency bands and amplitudes for which the microbeam escapes requires models able to follow the response up to the onset of instability.

\citet{nayfeh2007dynamic} identified three mechanisms of dynamic pull-in. The first is a bifurcation of the periodic orbit leading directly to pull-in, which does not depend on the initial conditions and gives an upper bound of the stable motions. In the second, a change of the excitation drives the response beyond a cyclic fold of the lower branch, and whether the microbeam settles on another orbit or pulls-in depends on the transient phase. In the third, the stable and unstable manifolds of the saddle intersect transversally as the excitation increases, the basin of bounded motions erodes, and small transient deviations from a stable periodic orbit can lead to pull-in. The last two mechanisms are governed by the global dynamics, which they noted becomes difficult to map when the reduced-order model has several modes and the phase space several dimensions. Under realistic disturbances, an attractor disappears before its bifurcation once its basin of attraction becomes too small or too fragmented \citep{Soliman_Thompson_1989,Rega_Lenci_2005}. Integrity measures quantify this erosion \citep{Soliman_Thompson_1989,Lenci_Rega_2003}, and their application to electrostatic microbeams has shown that the experimental limits of the pull-in bands coincide with given levels of erosion of the safe basin \citep{alsaleem2010experimental}, and that the observed range of existence of each branch can be predicted from its integrity \citep{ruzziconi2013electrically}. The erosion of the basin of electrostatic microactuators by fractal tongues, which renders the outcome highly sensitive to the initial conditions, was analysed by \citet{najar2010dynamics}. These analyses relied on structural models in which the dissipation is represented by a linear or lumped damping term. The effect of the dynamics of the film on the global stability of the microbeam has not been examined. A reduced-order model with few degrees of freedom that retains this dynamics is a prerequisite for such an analysis.

Unless the device is encapsulated in vacuum, the gas trapped between the microbeam and the rigid electrode governs largely contribute to its damping mechanism. The squeeze-film effect is described by the Reynolds equation, which couples the pressure in the gap to the motion of the microbeam \citep{hamrock2004fundamentals}. The film not only induce damping, but also stiffens the structure. Its elastic part becoming dominant when the gas cannot escape within a period of oscillation, and the balance between the two depends on the frequency, the gap and the ambient pressure \cite{sumali2007squeeze}. Most models replace the film by damping coefficients derived from the linearized Reynolds equation, either analytically or in terms of the structural mode shapes \citep{nayfeh2004squeeze,younis2007simulation,Amster2024}, or include it as a nonlinear analytical damping force in a structural reduced-order model \citep{khater2011dynamic}. Such partially coupled models have been shown to lose their accuracy near the pull-in instability and under resonant AC excitation, where the fully coupled fluid--structure problem must be solved \citep{najar2020multifidelity}. Fully coupled models, which solve the nonlinear Reynolds equation together with the structural equation, have been used to study the shock response \citep{Yagubizade_Younis_2011}, the response of microswitches \citep{sassi2018square}, capacitive microplates \citep{jallouli2019modeling} and arched microbeams \citep{Najar_Ghommem_Abdelkefi_2020}. In these models, however, the structure is usually reduced to a few real modes while the pressure field is resolved on a grid, so that the size of the problem is dominated by the fluid. More generally, the importance of nonlinear damping in the strongly nonlinear regimes of electromechanical resonators has recently been emphasized \citep{Gusso_Ujevic_2025}.

The linearized coupled problem is non-self-adjoint, since the film loads the microbeam through its pressure whereas the microbeam drives the film through its velocity. Its eigensolutions are therefore complex, and the coupled modes carry a phase difference between the deflection and the pressure that no real mode can represent \citep{jallouli2019modeling,Najar_Ghommem_Abdelkefi_2020}. A reduced-order model built on these complex modes, through a Petrov--Galerkin projection on the left eigenvectors, was proposed by \citet{najar2026complex}. Its spectrum contains, in addition to the coupled modes, real eigenvalues that describe the relaxation of the pressure field. It was shown that one of these fluid relaxation modes is indispensable for reproducing the forced response, and that adding coupled modes cannot replace it. That study was, however, restricted to transient responses at a given operating point.

In the present work, the reduced-order model based on the complex modes of the coupled problem is extended to the computation of periodic responses, their stability and their global stability, for clamped--clamped microbeams. Periodic orbits are computed by finite differences in time and continued in frequency, and their stability is determined from their Floquet multipliers \citep{sassi2018strong}. The model is first validated against the measured and computed frequency response of a microswitch \citep{sassi2018square}. Closed-form expressions are then derived for the fluid relaxation modes, which define a parameter characterizing the regime of the film and its correction under strong coupling. The approach is finally applied to a polysilicon microbeam fabricated and characterized in vacuum by \citet{tilmans1994electrostatically}, whose dynamics without film were studied in detail by \citet{najar2010nonlinear,najar2010dynamics}, operated here in a regime where the fluid modes relax on the timescale of the oscillation. For this microbeam, the effect of the fluid mode is analyzed on the frequency response and its bifurcations, on the basins of attraction through local and global integrity measures, and on dynamic pull-in maps in the plane of the excitation parameters.

The paper is organized as follows. Section \ref{sec:formulation} recalls the coupled model, its discretization, the complex-mode reduced-order model and the computation of periodic responses, and derives the fluid relaxation modes. Section \ref{sec:validation} validates the model against the microswitch. Section \ref{sec:dynamics} analyses the frequency response, the basins of attraction and the dynamic pull-in maps of the clamped--clamped microbeam. Conclusions are drawn in Section \ref{sec:conclusions}.

\section{Problem formulation and reduced-order model}\label{sec:formulation}
This section recalls the coupled model of an electrostatically actuated microbeam subjected to squeeze-film damping and the reduced-order model built on the complex modes of the coupled problem. The derivation of the eigenvalue problem and of the projection is given in detail in \citet{najar2026complex}, and only the elements needed in the present work are reproduced here. Two features are specific to the present study: the electrode may cover only part of the microbeam, and the reduced model is used to compute periodic responses over a range of excitation frequencies rather than single transient responses.

\subsection{Governing equations}
\begin{figure}
	\includegraphics[width=0.95\columnwidth]{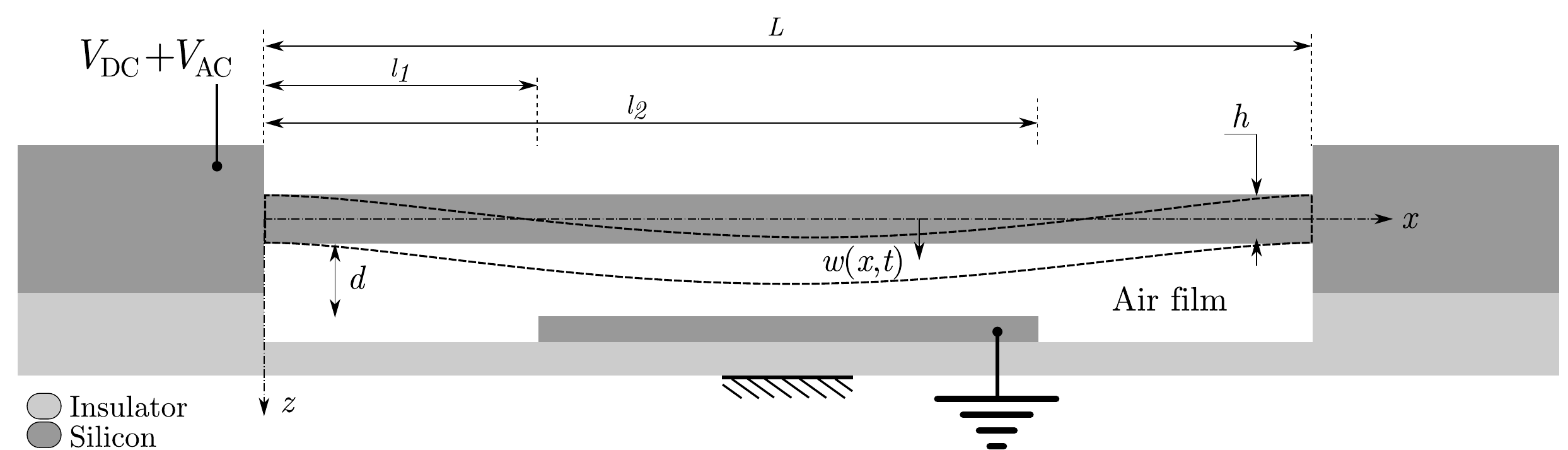}
	\caption{Schematic of a clamped--clamped microbeam with partially covered electrodes.} \label{fig1}
\end{figure}

A microbeam of length $L$, width $b$ and thickness $h$ is suspended at a distance $d$ above a substrate electrode, which extends over only part of its length, as shown in Figure \ref{fig1}. The microbeam is actuated by a voltage composed of a DC component and a harmonic AC component. The microbeam is modelled as an Euler--Bernoulli beam, including mid-plane stretching when both ends are clamped, residual axial stresses and the nonlinear dependence of the electrostatic force on the gap with a fringing-field correction. The air film between the microbeam and the electrode is described by the isothermal compressible Reynolds equation in two dimensions \citep{hamrock2004fundamentals}, which presumes $d \ll L,\,b$, laminar flow and negligible gas inertia, rarefaction is accounted for through an effective viscosity.

The nondimensional transverse deflection $w(x,t)$ and relative pressure $p(x,y,t)$ are governed by \citep{sassi2018square,najar2026complex}
\begin{align} 
	&\ddot w + \gamma\dot w + {w_{,xxxx}} = w_{,xx}  \left( {{N} - {\alpha _1}\int_0^1 w\,w_{,xx}dx}  \right)  \label{eq1} \\
	&\hspace{2.8cm} + {\alpha _2} S(x) \frac{{{{\left( {{V_{DC}} + V_{AC}(t)} \right)}^2}}}{{{{\left( {1 - w} \right)}^2}}}\left( {1 + \alpha _3 (1 - w) } \right) -{\alpha _4}\int_0^1 {p\, dy} \nonumber \\
	&\left[ {{(1-w)^3} (1+p)  p_{,x}} \right]_{,x} + \alpha_5 \left[ {{(1-w)^3} (1+p) p_{,y}} \right]_{,y} = \sigma \Big((1-w) \dot{p} - (1+p)\dot{w} \Big) \label{eq2}
\end{align}
subjected to the nondimensional boundary conditions
\begin{align}
	&w(0,t) =0, \; w(1,t) = 0, \; w_{,x} (0,t) =0 \mbox{ and }  w_{,x}(1,t) = 0 \label{eq3} \\
	&p(x,0,t) =0,\; p(x,1,t) =0,\;  p_{,x}(0,y,t) =0 \mbox{ and } p_{,x}(1,y,t) = 0 \label{eq4}
\end{align}
where the over-dots denote partial derivatives with respect to the nondimensional time, the subscripts $x$ and $y$ after a comma denote spatial partial derivatives, and $V_{AC}(t) = v_{ac}\cos(\Omega t)$, $\Omega$ is the nondimensional excitation frequency. The film is vented to ambient pressure along its two long edges and closed at the clamped ends. The mid-plane stretching term is written in the form $-\int_0^1 w\,w_{,xx}\,dx$, obtained from $\int_0^1 w_{,x}^2\,dx$ by integration by parts, which is permissible since the deflection vanishes at both clamped ends.

The function $S(x)$ in Equation \eqref{eq1} describes the extent of the electrode. A sharp electrode edge would introduce a discontinuity in the electrostatic load that the differential quadrature approximation resolves poorly. The electrode is therefore represented by the smooth window
\begin{equation} \label{eq5}
	S(x) = \frac{1}{2}\left[\tanh\left(\frac{x-l_1}{\delta}\right) - \tanh\left(\frac{x-l_2}{\delta}\right)\right]
\end{equation}
where $l_1$ and $l_2$ are the nondimensional positions of the electrode edges and $\delta$ controls the width of the transition. For $\delta \ll l_2-l_1$ the integral of $S$ equals $l_2-l_1$, so that the window preserves the electrode area. A microbeam covered by the electrode over its whole length is recovered with $S(x)=1$.

The structural damping coefficient $\gamma$ represents the material and anchor losses, assumed viscous and uniform, whereas the effect of the gas film, both dissipative and elastic, enters the structural equation exclusively through the term $\alpha_4 \int_0^1 p\,dy$. The nondimensional variables and parameters are
\begin{equation}  \label{eq6}
	\begin{gathered}
		x = \frac{{\hat x}}{L},\,\,y = \frac{{\hat y}}{b},\,\,w = \frac{{\hat w}}{d},\,\, t = \frac{{\hat t}}{\tau} , \,\, p= \frac{{\hat p-p_a}}{p_a}, \,\, \tau= \sqrt{\frac{{12 \rho {L^4}}}{{Eh^2}}},  \,\, N = \frac{{\hat N{L^2}}}{{EI}}, \,\, \gamma = \frac{\hat c L^4}{EI\,\tau} \\
		{\alpha _1} = 6\frac{{{d^2}}}{{{h^2}}}, \,\, {\alpha _2} = \frac{{6\varepsilon {L^4}}}{{E{h^3}{d^3}}}, \,\,  {\alpha _3} = \frac{{0.65}}{{b/d}}, \,\, {\alpha _4} =\frac{12  p_a L^4 }{d E h^3} , \,\, {\alpha _5} = \frac{L^2}{b^2} , \,\,  \sigma=\frac{12 \mu_{eff} L^2}{\tau p_a d^2}
	\end{gathered}
\end{equation}
where hats denote corresponding dimensional quantities, $E$ and $\rho$ are the Young's modulus and density of the microbeam, $I = bh^3/12$ is the second moment of area of its cross-section, $\varepsilon$ is the permittivity of the gap medium, $\hat c$ is the structural damping coefficient per unit length, $\hat N$ is the resultant axial force arising from residual stresses and $p_a$ is the ambient pressure. The effective viscosity is taken as in \citet{sassi2018square}, following \citet{andrews1993comparison},
\begin{equation} \label{eq7}
	\mu _{eff}= \frac{\mu}{1+6.8636 \, {K_n}^{0.9906}}, \qquad K_n=\frac{\lambda_0 \, p_0 }{p_a d}
\end{equation}
where $\mu$ is the dynamic viscosity of air, $\lambda_0=65 \times 10^{-9}$m is the mean free path at the standard pressure $p_0=1.01325 \times 10^5\,$Pa and $K_n$ is the Knudsen number.

The two fields are coupled asymmetrically since the microbeam is loaded by the pressure through the term $\alpha_4 \int_0^1 p\,dy$, whereas the film is driven by the velocity of the microbeam through the term $\sigma(1+p)\dot w$. The linearized operator is therefore non-self-adjoint, its eigenvalues are in general complex, and its left and right eigenvectors are distinct. This property underlies the reduced basis of Section \ref{sec:rom}.

\subsection{Discretization and coupled eigenvalue problem}\label{sec:evp}
Equations \eqref{eq1} and \eqref{eq2} are discretized following \citet{najar2026complex}. The structural equation and its boundary conditions are discretized with the locally adaptive differential quadrature method (LaDQM) \citep{ng2009accuracy}, one fictitious point is added beyond each clamped end to carry the boundary conditions so that every interior node remains available for coupling with the film and the electrostatic force. The Reynolds equation is discretized with the classical differential quadrature method (DQM) \citep{bellman1971differential,bert1996differential} on a two-dimensional grid. Both domains share the same Chebyshev--Gauss--Lobatto grid of $n$ points along $x$, which avoids any interpolation in the coupling terms, and the same distribution is used along $y$:
$$x_i=\frac12\left( 1-\cos \frac{i-1}{n-1}\pi\right) \mbox{ and } y_i=\frac12\left( 1-\cos \frac{i-1}{n-1}\pi\right)$$
The pressure field is symmetric about $y=\tfrac12$ for any deflection, so that only the nodes $j = 2, \dots, \frac{n+1}{2}$ are retained across the width, whereas no symmetry is imposed along $x$. The integral terms are evaluated with the quadrature weights $T_i$ obtained by integrating the Lagrange polynomials on which the differential quadrature is built, given by Equation \eqref{eqNC}.

The static deflection $u(x)$ satisfies Equations \eqref{eq1} and \eqref{eq3} with the time derivatives and the AC component suppressed, the relative pressure vanishing at equilibrium. It is computed with a Newton--Raphson procedure \citep{najar2005modeling}. In the presence of an AC component, the static solution is computed at the root-mean-square voltage $V_{RMS}=\sqrt{V_{DC}^2+\frac12 v_{ac}^2}$ rather than at $V_{DC}$, which places the linearization about the mean operating point of the response.

Writing $w(x,t)=u(x)+q(x,t)$ and linearizing about $u$ and $p=0$, the discretized equations take the form
\begin{equation} \label{eq8}
	\bf M \ddot x +  C \dot x + K(u) x =  0, \; {\bf x}= \left( \begin{array}{c} {\bf{q}}  \\ \bf p  \end{array}   \right)
\end{equation}
where ${\bf q}$ collects the $n_w = n-2$ interior deflections and ${\bf p}$ the $n_p = (n-2)\frac{n-1}{2}$ retained pressures. The mass matrix $\bf M$ is singular since the Reynolds equation is only of first order in time. Also, the stiffness and damping matrices are non-symmetric and have their block structure given by 
\begin{equation} \label{eq9}
	{\bf K}= \left[ \begin{array}{cc}
		{\bf K}_{11} & {\bf K}_{12}\\
		{\bf 0} & {\bf K}_{22}
	\end{array}\right], \qquad
	{\bf C}= \left[ \begin{array}{cc}
		{\bf C}_{11}&{\bf 0} \\
		{\bf C}_{21}&{\bf C}_{22}
	\end{array}\right]
\end{equation}
in which the pressure loads the structure through ${\bf K}_{12}$ and the structure drives the pressure through ${\bf C}_{21}$, with ${\bf C}_{11}=\gamma{\bf I}$ and ${\bf C}_{22}$ diagonal with entries proportional to $\sigma(1-u_i)$. Writing Equation \eqref{eq8} in state space and eliminating the pressure-rate unknowns, which do not appear in the mass matrix, yields the classical eigenvalue problem \citep{najar2026complex}
\begin{equation} \label{eq10}
	{\bf \hat B \Phi } = \lambda {\bf \Phi}, \qquad
	{\bf \hat B}= \left[ \begin{array}{ccc}
		{\bf 0} & {\bf 0} & {\bf I} \\
		{\bf 0} & -{\bf C}_{22}^{-1} {\bf K}_{22} & -{\bf C}_{22}^{-1} {\bf C}_{21} \\
		-{\bf K}_{11} & -{\bf K}_{12} & -{\bf C}_{11}
	\end{array}\right], \qquad
	{\bf \Phi} = \left( \begin{array}{c} {\bf \Phi^1} \\ {\bf \Phi^2} \\ {\bf \Phi^3} \end{array}  \right)
\end{equation}
of dimension $2n_w+n_p$, where ${\bf \Phi^1}$, ${\bf \Phi^2}$ and ${\bf \Phi^3}$ are the displacement, pressure, and velocity partitions of the eigenvector. Since ${\bf \hat B}$ is real and non-symmetric, its right eigenvectors are not mutually orthogonal, and the adjoint problem
\begin{equation} \label{eq11}
	{\bf \hat B} ^T {\bf \Psi} = \lambda {\bf \Psi}
\end{equation}
is introduced. The left eigenvectors ${\bf \Psi}$, partitioned as ${\bf \Phi}$, are scaled so that the two sets satisfy the biorthonormality conditions
\begin{equation} \label{eq12}
	{\bf \Psi}_s^T {\bf \Phi}_r = \delta_{rs}, \qquad
	{\bf \Psi}_s^T {\bf \hat B} {\bf \Phi}_r = \lambda_r \delta_{rs}
\end{equation}

The eigenvalues of ${\bf \hat B}$ contain two sets of different nature. A first set with complex-conjugate pairs describes coupled oscillations of the microbeam and the air film, characterized by the undamped frequency, damping ratio, and quality factor, given by
\begin{equation} \label{eq13}
	\omega_r = \left| \lambda_r \right|, \qquad
	\xi_r= \frac{-\mbox{Re}(\lambda_r)}{\left| \lambda_r \right| }, \qquad
	Q_r=\frac{1}{2 \xi_r}.
\end{equation}
The second set is formed by real negative eigenvalues depicting the non-oscillatory relaxations of the pressure field, which have no counterpart in the undamped structural problem. For these fluid modes $Q_r=\tfrac12$, and they are characterized instead by their relaxation time $\tau_r = 1/|\lambda_r|$.

\subsection{Fluid relaxation modes}\label{sec:fluidmodes}
When the film is weakly coupled to the microbeam, the fluid relaxation modes have a negligible structural content, and their eigenvalues are governed by the linearized Reynolds equation alone. When the gap is uniform, as at zero bias, this equation reduces to
\begin{equation} \label{eqF1}
	p_{,xx} + \alpha_5\, p_{,yy} = \sigma \nu \, p
\end{equation}
which, with the boundary conditions \eqref{eq4}, admits the solutions
\begin{equation} \label{eqF2}
	p_{rs}(x,y) = \cos(r\pi x) \, \sin(s\pi y)
\end{equation}
with the eigenvalues
\begin{equation} \label{eqF3}
	\nu_{rs} = -\frac{\pi^2}{\sigma} \left(r^2 + \alpha_5\, s^2\right) , \qquad r = 0,1,2,\dots, \quad s = 1,2,\dots
\end{equation}
Since the pressure field is symmetric about $y=1/2$, only odd values of $s$ occur, and even and odd values of $r$ give the modes that are symmetric and antisymmetric along $x$, respectively. For a long and narrow film, $\alpha_5 \gg 1$, the lowest fluid modes all have $s=1$ and differ only through their variation along the microbeam, the slowest one is $\nu_{01} = -\pi^2\alpha_5/\sigma$. Under a uniform static deflection $u$, the eigenvalues \eqref{eqF3} are multiplied by $(1-u)^2$.

The regime of the film is set by the rate at which its slowest mode relaxes compared with the motion of the microbeam. This mode responds quasi-statically to the microbeam motion when its eigenvalue is much larger than the natural frequency of the structure, so that a ratio $\Lambda$ of this eigenvalue to the first natural frequency $\omega_1$ characterizes the film. For $\Lambda \gg 1$ the fluid modes act as quasi-static corrections to the pressure distribution, whereas for $\Lambda$ of order one or smaller the slowest fluid mode relaxes on the timescale of the oscillation and carries its own dynamics. In the weak-coupling limit, one can calculate $\Lambda$ as follows:
\begin{equation} \label{eqF4}
	\Lambda = \frac{|\nu_{01}|}{\omega_1} = \frac{\pi^2 \alpha_5}{\sigma\,\omega_1}
\end{equation}

Equation \eqref{eqF4} neglects the structural content of the fluid modes and is therefore restricted to films weakly coupled to the microbeam. When the coupling is strong, the fluid eigenvalues cannot be represented by Equation \eqref{eqF3}, and $\Lambda$ is evaluated directly from the computed spectrum, as shown in Section \ref{sec:dynamics}.

\subsection{Reduced-order model based on the complex modes}\label{sec:rom}
Decomposing the deflection as $w=u+q$, Equations \eqref{eq1} and \eqref{eq2} are written in state-space form:
\begin{align}
	&\dot{z}_1 = z_3 \label{eq14} \\
	&\dot{z}_2 = -\frac{3}{\sigma} (u_{,x} +z_{1,x}) (1- u -z_1) (1+z_2) z_{2,x} \label{eq15} \\
	& \hspace{0.8cm} + \frac{1}{\sigma} (1- u -z_1)^2 (z_{2,x}^2 + z_{2,xx} + z_2 z_{2,xx} + \alpha_5 z_{2,y}^2 + \alpha_5 z_{2,yy} + \alpha_5 z_2 z_{2,yy} )   \nonumber\\
	&  \hspace{0.8cm} + \frac{(1+z_2)z_3}{1- u - z_1} \nonumber\\
	&\dot{z}_3 = - \gamma z_3 - {u_{,xxxx}} - {z_{1,xxxx}}  \label{eq16} \\
	& \hspace{0.8cm}+ (u_{,xx} + z_{1,xx} )  \left( {{N} - {\alpha _1} \int_0^1 (u+z_1)(u_{,xx}+z_{1,xx}) \, dx}  \right) \nonumber \\
	& \hspace{0.8cm}  + {\alpha _2} S(x) \frac{{{{\left( {{V_{DC}} + V_{AC}(t)} \right)}^2}}}{{{{\left( {1 - u- z_1} \right)}^2}}}\left( {1 + \alpha _3 (1 - u -z_1) } \right) - {\alpha _4}\int_0^1 {z_2\, dy} \nonumber
\end{align}
where $z_1=q(x,t)$, $z_2=p(x,y,t)$ and $z_3=\dot{q}(x,t)$. The vector ${\bf z}$ collecting the values of $z_1$, $z_2$ and $z_3$ at the DQM grid points is expanded on the right eigenvectors of ${\bf \hat B}$,
\begin{equation}\label{eq17}
	{\bf z} = \left( \begin{array}{c} {\bf z^1} \\ {\bf z^2} \\ {\bf z^3} \end{array}\right)
	= \sum_{r \in \mathcal{E}_C} 2\,\mathrm{Re} \left( \eta_{r}(t)\, {\bf \Phi}_r\right)  + \sum_{r \in \mathcal{E}_F} \eta_{r}(t)\, {\bf \Phi}_r
\end{equation}
where $\mathcal{E}_C$ is the set of retained coupled modes where $\mathrm{Im}(\lambda_r)>0$, and $\mathcal{E}_F$ the set of retained fluid modes. The coordinates $\eta_r$ are complex for $r\in\mathcal{E}_C$ and real for $r\in\mathcal{E}_F$, and taking twice the real part reconstitutes the contribution of the conjugate modes which keeps ${\bf z}$ real.

Writing the right-hand side of Equations \eqref{eq14}--\eqref{eq16} as ${\bf \hat F}({\bf z},t) = {\bf \hat B}{\bf z} + {\bf \hat N}({\bf z},t)$, substituting Equation \eqref{eq17} and projecting the residual onto the left eigenvectors, and using the biorthonormality conditions \eqref{eq12}, yields,
\begin{equation}\label{eq18}
	\dot{\eta}_s = \lambda_s {\eta}_s + {{\bf \Psi}^{\bf{2}}_s}^T {\bf\hat N^2}\left( {\bf z} \right)  + {{\bf \Psi}^{\bf{3}}_s}^T {\bf\hat N^3}\left( {\bf z} \right), \quad s \in \mathcal{E}_C \cup \mathcal{E}_F
\end{equation}
The first partition of ${\bf \hat N}$ vanishes, since Equation \eqref{eq14} is linear, and the two others are given by
\begin{align}
	{\bf\hat N^2} \left( {\bf z} \right) = &-\frac{3}{\sigma} {\bf E}  \left( {\bf a}^{(1)} {\bf u} + {\bf a}^{(1)} {\bf z^1} \right) \circ \left( {\bf EH} \right)   \circ ( {\bf 1} + {\bf z^2} ) \circ \left( {\bf b}_x^{(1)} {\bf z^2}\right)  \label{eq19} \\
	& + \frac{1}{\sigma} \Big[ \left( {\bf EH} \right)^{\circ 2} \circ \Big( \left( {\bf b}_x^{(1)} {\bf z^2} \right)  ^{\circ 2} + {\bf b}_x^{(2)} {\bf z^2} + {\bf z^2} \circ \left( {\bf b}_x^{(2)} {\bf z^2} \right) \nonumber \\
	&  + \alpha_5 \left( {\bf b}_y^{(1)} {\bf z^2} \right) ^{\circ 2} + \alpha_5 {\bf b}_y^{(2)} {\bf z^2} + \alpha_5 {\bf z^2} \circ \left( {\bf b}_y^{(2)} {\bf z^2} \right) \Big)   \Big] \nonumber\\
	& + ({\bf 1} + {\bf z^2}) \circ \left( {\bf E }{\bf z^3}\right)  \circ \left( {\bf EH} \right)^{\circ -1} + {\bf C}_{22}^{-1} {\bf K}_{22} {\bf z}^{\bf 2} + {\bf C}_{22}^{-1} {\bf C}_{21} {\bf z}^{\bf 3} \nonumber \\
	{\bf\hat N^3} \left( {\bf z} \right) = & - \gamma {\bf z^3} - {\bf a}^{(4)} {\bf u} - {\bf a}^{(4)} {\bf z^1}  \label{eq20} \\
	& + ( {\bf a}^{(2)} {\bf u} + {\bf a}^{(2)} {\bf z^1} )  \left( {N} - {\alpha _1} {\bf \tilde T}^T \left( ({\bf u} + {\bf z^1}) \circ ( {\bf a}^{(2)} {\bf u} +{\bf a}^{(2)} {\bf z^1} ) \right) \right) \nonumber \\
	& + {\alpha _2} \left( {{V_{DC}} + V_{AC}(t)} \right)^2 \left( {\bf H} \right)^{\circ -2} \circ \left( {\bf 1} + \alpha_3 {\bf H} \right) \circ {\bf S}  - {\alpha _4} {\bf G} {\bf z^2}  \nonumber \\
	& + {\bf K}_{11} {\bf z}^{\bf 1} + {\bf K}_{12} {\bf z}^{\bf 2} + {\bf C}_{11} {\bf z}^{\bf 3} \nonumber
\end{align}
where $\circ$ denotes the elementwise (Hadamard) product, ${\bf v}^{\circ k}$ the elementwise $k^{th}$ power of a vector ${\bf v}$, ${\bf H} = {\bf 1}- {\bf u} - {\bf z^1}$ is the gap, ${\bf S}$ collects the values of the electrode window $S(x_i)$ at the interior nodes, and $\tilde{\bf T} = (T_2, \dots, T_{n-1})^T$, the stretching integral involves the interior weights only since the deflection vanishes at both clamped ends. The transfer operators ${\bf E}$ and ${\bf G}$ and the condensed derivative matrices ${\bf a}^{(k)}$, ${\bf b}_x^{(k)}$ and ${\bf b}_y^{(k)}$ are defined in \ref{ApII}.

For numerical integration, Equation \eqref{eq18} is rewritten in real form by letting $\eta_r = a_r + \mathrm{i}\, b_r$ and $\lambda_r = \varsigma_r + \mathrm{i}\,\omega_{d,r}$, which gives
\begin{equation} \label{eq21}
	\dot a_r = \varsigma_r a_r - \omega_{d,r} b_r + \mathcal{R}_r, \qquad
	\dot b_r = \omega_{d,r} a_r + \varsigma_r b_r + \mathcal{I}_r
\end{equation}
where $\mathcal{R}_r$ and $\mathcal{I}_r$ are the real and imaginary parts of the projected terms in Equation \eqref{eq18}. Fluid modes have $\omega_{d,r}=0$ and $b_r \equiv 0$.

The electrostatic load is symmetric about $x=1/2$, only symmetric modes are therefore retained in the bases considered in this work. A basis is denoted C$m$F$l$, where $m$ is the number of symmetric coupled modes and $l$ the number of symmetric fluid modes, each set is taken in increasing order of $|\lambda_r|$. The corresponding reduced model comprises $2m+l$ real unknowns.

\subsection{Periodic responses}\label{sec:periodic}
Under the voltage $V_{DC}+v_{ac}\cos(\Omega t)$, Equations \eqref{eq21} form a non-autonomous system whose right-hand side is periodic with period $T_\Omega = 2\pi/\Omega$. Periodic responses are sought as solutions of Equations \eqref{eq21} satisfying the periodicity conditions
\begin{equation} \label{eq22}
	a_r(t+T_\Omega) = a_r(t), \qquad b_r(t+T_\Omega) = b_r(t), \qquad r \in \mathcal{E}_C \cup \mathcal{E}_F
\end{equation}

The resulting periodic boundary-value problem is discretized in time over one forcing period by 100 time nodes finite differences approximations \cite{nayfeh2008applied,najar2006dynamic}, and the periodic solutions are continued with respect to the excitation frequency $\Omega$ using a pseudo-arclength continuation procedure \citep{keller1977numerical,nayfeh2008applied}. The stability of each periodic solution is assessed from its Floquet multipliers computed by a shooting technique.

Following the measurements used for validation by \citet{sassi2018square}, the response is characterized by the root-mean-square velocity of the midspan over one period,
\begin{equation} \label{eq23}
	\dot w_{\mathrm{rms}}(\tfrac12) = \sqrt{\frac{1}{T_\Omega}\int_0^{T_\Omega} \dot w^2(\tfrac12,t)\,dt}
\end{equation}
where $\dot w(\tfrac12,t)$ is the component of ${\bf z^3}$ at the midspan node, $x_{(n+1)/2}=\tfrac12$, reconstructed from the modal coordinates through Equation \eqref{eq17}.

% ---------------------------------------------------------------------------
\section{Validation against a clamped--clamped microswitch} \label{sec:validation}

\subsection{Reference configuration} \label{sec:val-config}
The proposed reduced-order model is used to compute and validate the frequency-response curves of the capacitive shunt microswitch of \citet{sassi2018square}, for which curves were both computed and measured in ambient air. The switch is an electroplated gold bridge clamped at both ends and suspended over a coplanar waveguide, whose signal line acts as the actuation electrode. Its parameters are listed in Table \ref{tab:switch}. The squeeze film is vented to ambient pressure along the two long edges of the bridge and closed at the clamped ends. 

\begin{table}[H]
	\centering
	\caption{Parameters of the clamped--clamped microswitch \citep{sassi2018square}.} \label{tab:switch}
	\begin{tabular}{lll}
		\hline
		Parameter                     & Symbol      & Value                       \\ \hline
		Young's modulus               & $E$         & $80\,$GPa                   \\
		Density                       & $\rho$      & $19\,320\,$kg\,m$^{-3}$     \\
		Length                        & $L$         & $225\,\upmu$m               \\
		Width                         & $b$         & $20\,\upmu$m                \\
		Thickness                     & $h$         & $1.7\,\upmu$m               \\
		Initial gap                   & $d$         & $1.7\,\upmu$m               \\
		Axial load                    & $\hat N$    & $0.578\,$mN                 \\
		Coplanar waveguide dimensions & $G/W/G$     & $19/45/19\,\upmu$m          \\
		Electrode extent              & $x/L$       & $0.385$--$0.615$            \\
		Linear viscous damping        & $\gamma$    & $0.129$ (non-dimensional)   \\
		Ambient pressure              & $p_a$       & $101.325\,$kPa              \\
		Dynamic viscosity of air      & $\mu$       & $1.983\times10^{-5}\,$Pa\,s \\
		Mean free path at $p_0$       & $\lambda_0$ & $65\,$nm                    \\ \hline
	\end{tabular}
\end{table}

For this microswitch, to be consistent with the assumptions taken in \citep{sassi2018square}, gas rarefaction is accounted for through the effective viscosity
\begin{equation}\label{eq:mueff}
	\mu_{\mathrm{eff}} = \frac{\mu}{1 + 6.8636\,K_n^{0.9906}}
\end{equation}

The reference model couples the Euler--Bernoulli beam, including mid-plane stretching, with the two-dimensional nonlinear Reynolds equation. Its dynamic response was computed with a hybrid reduced model with a single Galerkin mode of the unactuated bridge for the structure, and a $49\times49$ differential-quadrature grid for the pressure. Periodic orbits were obtained by finite differences. The fluid is therefore described by the pressures at the interior grid nodes, while the structure is restricted to one real mode computed without the fluid coupling.

The present model uses similar assumptions, however, the structure is no longer restricted to a mode of the unactuated bridge, and the pressure is no longer resolved on the full grid. Both fields are instead projected on the complex modes of the coupled problem linearized about the static deflection. Only symmetric modes are retained, since the antisymmetric ones have no projection on the symmetric electrostatic load. The electrode is represented by the nondimensional smooth window given by Equation \eqref{eq5}, where $ l_1 = 0.385$, $l_2 = 0.615$, and $\delta = 0.025$. The linear viscous damping $\gamma$, which accounts for the structural losses, is chosen so that the corresponding quality factor is $Q_v = 250$ estimated by \citet{khater2011dynamic}. It is given by $\gamma = 0.129$.

% ---------------------------------------------------------------------------
\subsection{Static and linear response} \label{sec:val-linear}
The static pull-in voltage obtained with the present discretization is compared in Table \ref{tab:pullin} with the DQM, finite-element and experimental values of \citet{sassi2018square}.

\begin{table}[H]
	\centering
	\caption{Static pull-in voltage.}
	\label{tab:pullin}
	\begin{tabular}{lcccc}
		\hline
		                      & Present &  DQM   &  FEM   & Experiment \\
		$V_{\mathrm{PI}}$ (V) & $64.5$  & $64.2$ & $63.2$ &   $64.5$   \\ \hline
	\end{tabular}
\end{table}

\begin{table}[H]
	\centering
	\caption{Lowest symmetric eigenvalues of the linearized coupled problem. Frequency $f$ and quality factor $Q$ for the coupled modes denoted by C, relaxation time $\tau$ for the fluid modes denoted by F.}
	\label{tab:spectrum}
	\begin{tabular}{lcccc}
		\hline
		      		  &   \multicolumn{2}{c}{$(V_{\mathrm{dc}}, v_{\mathrm{ac}}) = (0, 0)\,$V}  & \multicolumn{2}{c}{$(V_{\mathrm{dc}}, v_{\mathrm{ac}}) = (29, 28.6)\,$V} \\
				Mode  & $f$ (kHz) &   $Q$    			& $f$ (kHz) &   $Q$    \\ \hline
				1st C & $101.19$  & $15.54$  			& $95.89$   & $12.94$  \\
				2nd C & $433.23$  & $70.86$  			& $432.70$  & $64.87$  \\
				3rd C & $997.70$  & $182.84$ 			& $997.65$  & $168.83$ \\ \hline
					  & \multicolumn{2}{c}{$t_r$ (ns)} 	& \multicolumn{2}{c}{$t_r$ (ns)} \\ \hline
				1st F & \multicolumn{2}{c}{$25.94$} 	& \multicolumn{2}{c}{$28.24$} \\
				2nd F & \multicolumn{2}{c}{$25.15$} 	& \multicolumn{2}{c}{$26.15$} \\
				3rd F & \multicolumn{2}{c}{$23.03$} 	& \multicolumn{2}{c}{$24.30$} \\ \hline
	\end{tabular}
\end{table}

The coupled problem is discretized with $n = 27$ DQM points in each direction for the microbeam and the film. Table \ref{tab:spectrum} lists the lowest symmetric eigenvalues of the linearized coupled problem at zero bias and at the operating point of the frequency-response curves, $(V_{\mathrm{dc}}, v_{\mathrm{ac}}) = (29, 28.6)\,$V. At this point the problem is linearized about the static deflection under the RMS voltage $V_{\mathrm{rms}} = \sqrt{V_{\mathrm{dc}}^2 + \tfrac12 v_{\mathrm{ac}}^2} = 35.36\,$V. These eigenvalues are those of the full discretized problem and do not depend on the reduced basis. The time responses of this section, namely the simulated ring-down and the frequency-response curves, are computed with the reduced model of Section \ref{sec:rom}, using the C1F1 basis, composed of the first symmetric coupled mode and the slowest symmetric fluid mode, unless another basis is indicated.

At zero bias, the first coupled mode has a frequency of $101.19\,$kHz which is $0.2$\% higher than the measured $101\,$kHz, and $0.05$\% higher than the $101.14\,$kHz of the reference model \citep{sassi2018square}. Its quality factor is $15.54$, $20$\% above the measured value $Q = 13.0$. Without structural damping the model gives $Q_s = 16.7$ against $13.7$ deduced from the measurement, so that the difference lies in the film model, which underestimates the squeeze-film dissipation by $20$\%. In fact, at the Knudsen number of the switch ($K_n = 0.038$), the expression used in the reference model \citep{sassi2018square} and those of \citet{Veijola1995}, \citet{burgdorfer1959influence} and \citet{hsia1983investigation} gives $\mu_{eff}/\mu$ between $0.79$ and $0.82$, about $4$\% difference. The difference between these models is five times smaller than the discrepancy with the measured number. Moreover, matching the measurement would require the rarefaction to reduce the viscosity by only $6$\%, against $18$ to $21$\% for these models, a correction about four times weaker.

A more plausible cause is the flow beyond the open edges. The boundary conditions $p=0$ at $y=0$ and $y=1$ assume that the pressure returns to its ambient value exactly at the edges of the microbeam, whereas the gas escaping laterally still meets a viscous resistance beyond them. This border effect is known to make the Reynolds equation underestimate the squeeze-film damping \citep{mol2009squeezed,lu2021trt}, and it is commonly accounted for by extending the surface dimensions by a length of the order of the gap \citep{veijola2005extending}. Increasing the width of the film by $1.3\,\upmu$m, that is $0.76\,d$, reproduces $Q_s = 13.7$. However, it moves the upper branch of the frequency response further below the measurements near resonance. The remaining discrepancy on the large orbits is therefore not a matter of linear damping. The film model is kept identical to that of the reference so that the comparison is only related to the model reduction.

Similarly to the measurement procedure, a ring-down test simulated with the reduced model C1F1 at $0.4\,$V square pulse of $10\,\upmu$s was applied to the switch at rest, and the same logarithmic-decrement fit is applied to the first nine peaks of the free decay of the midspan velocity. The fit gives $Q = 15.49$, $0.3$\% of the eigenvalue. Therefore, the C1F1 basis is adequate to completely recover the linear decay.

Under bias, the frequency of the first mode drops to $95.89\,$kHz and its quality factor to $12.94$ as the static deflection narrows the gap and increases the dissipation in the film. At this operating point, the first symmetric coupled mode and its complex conjugate are followed by ten coupled pairs, five symmetric and five antisymmetric. A basis selected by increasing frequency would therefore contain at least 24 coupled states before reaching any fluid mode. The slowest fluid mode relaxes in $25.9$\,ns at zero bias and $28.2$\,ns under bias. Near fundamental resonance, its eigenvalue is about $59$ times the excitation frequency, so all the fluid modes respond quasi-statically.

% ---------------------------------------------------------------------------
\subsection{Frequency-response curves}
\label{sec:val-frc}
Periodic orbits are computed by finite differences in time and continued in frequency by pseudo-arclength continuation \citep{nayfeh2008applied,najar2006dynamic}. Stability of the orbits and the end points of the branches are assessed by calculating the Floquet multipliers over the continuation scheme using the shooting method. Following the measurements, the response is characterized by the RMS velocity of the midspan over one period, $\dot w_{\mathrm{rms}}(\tfrac12)$. The harmonic waveform $V(t) = V_{\mathrm{dc}} + v_{\mathrm{ac}}\cos\Omega t$ is applied with $(V_{\mathrm{dc}}, v_{\mathrm{ac}}) = (29, 28.6)\,$V \citep{sassi2018square}.

\begin{figure}
	\centering
	\includegraphics[width=0.6\columnwidth]{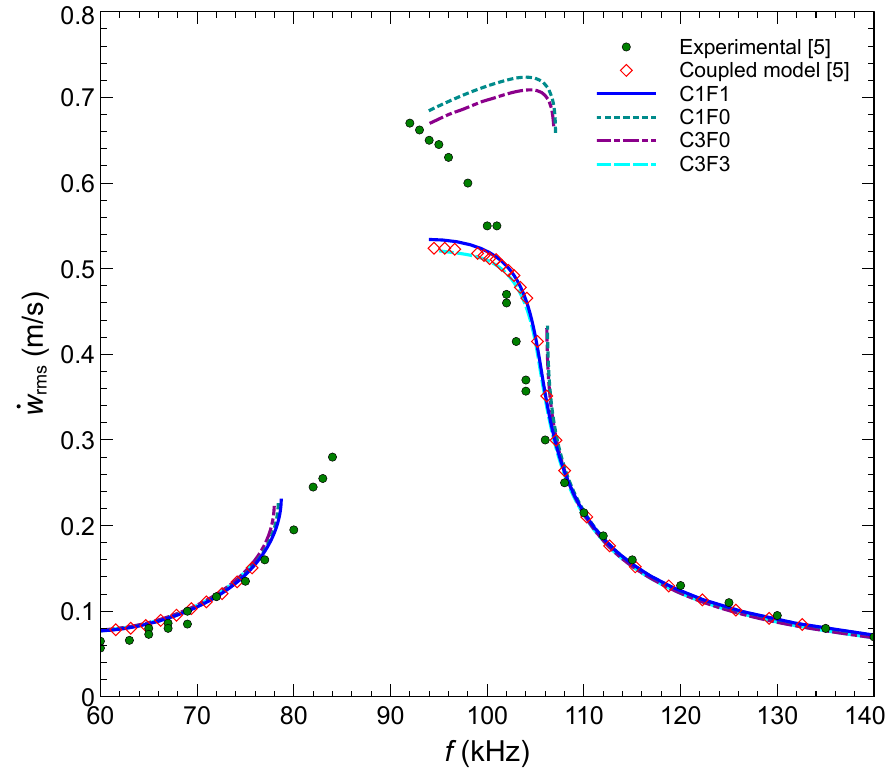}
	\caption{Stable orbits of the midspan RMS velocity of the frequency-response curves for $(V_{\mathrm{dc}}, V_{\mathrm{ac}}) = (29, 28.6)\,$V. Dots: experiment, diamonds: coupled model \citep{sassi2018square}, and lines: present reduced models.} \label{fig:val-frc}
\end{figure}

Figure \ref{fig:val-frc} compares the measured response and the reference coupled model with four reduced models: C1F0, C3F0, C1F1 and C3F3. C$m$F$l$ denotes a basis of $m$ coupled modes and $l$ fluid modes, while all selected modes are symmetric. Only stable solutions are shown in this figure.

In Figure \ref{fig:val-frc}, the C1F1 model with three real degrees of freedom, follows the reference coupled model on both branches over the whole range where the reference was computed. The reference describes the film by $2209$ pressure unknowns. The C3F3 basis differs from C1F1 only on the upper branch, where it is slightly below it, which indicates that the C1F1 basis is close to convergence.

Without fluid modes, the small and large orbits are affected differently. On the lower branch, where the motion is small, the C$m$F0 curves remain close to C1F1 and their fold is barely shifted. On the upper branch, the C1F0 model overestimates the response and shifts the curves to slightly higher frequencies, introducing a couple of bifurcation points at large orbits. Adding coupled modes does not correct this, as shown by the difference between C3F0 and C1F0. The missing contribution therefore lies in the fluid set of the spectrum, and it matters only at large amplitude. The fluid mode, which responds quasi-statically, supplies correction, and its omission leads to an underestimate of the dissipation in the film.

On the upper branch, all the converged models remain below the measured velocities. This confirms the observation of \citet{sassi2018square} for $\dot w_{\mathrm{rms}} > 0.45\,$m/s, which the reference attributed either to the limits of the Reynolds equation at small gaps or to higher modes absent from its single-mode basis. As C3F3 continues to deviate from the measurements and remains close to C1F1, modal truncation is disregarded, and the discrepancy is attributed to the film model.

%%%%%%%%%%%%%%%%%%%%%%%%%%%%%%%%%%%%%%%%%%%%%%%%%%%%%%%%%%%%%%%%%%%%%%%%%%%%%%%%%%%%%%%%%%%%%%%%%%%%%%%%%%%%%%%%%%%%%%%%%%%%%
\section{Dynamic response and global stability}\label{sec:dynamics}
The reduced-order model is now applied to a clamped--clamped polysilicon microbeam from the vacuum-encapsulated resonators fabricated and characterized by \citet{tilmans1994electrostatically}. The nonlinear dynamics of this microbeam were studied in detail by \citet{najar2010nonlinear,najar2010dynamics}, who reported locally hardening and softening behaviors, cyclic-fold and period-doubling bifurcations, dynamic pull-in and the erosion of the basin of attraction of bounded motions. These studies neglected the squeeze film and discretized the microbeam by the differential quadrature method with $9$ and $5$ grid points, respectively.

The microbeam parameters are listed in Table \ref{tab:beam} the elastic modulus is identified by \citet{tilmans1994electrostatically} from the measured resonance frequencies. Since these measurements were made in vacuum, they characterize the microbeam without film, and the results obtained below in the presence of air are predictions of the model. For this microbeam, the electrode covers the whole length $S(x)=1$.

\begin{table}[htbp]
	\centering
	\caption{Parameters of the clamped--clamped microbeam \citep{tilmans1994electrostatically,najar2010nonlinear}.}
	\label{tab:beam}
	\begin{tabular}{lcc}
		\hline
		Parameter                 & Symbol   & Value                      \\ \hline
		Length                    & $L$      & $510\,\upmu$m              \\
		Width                     & $b$      & $100\,\upmu$m              \\
		Thickness                 & $h$      & $1.5\,\upmu$m              \\
		Initial gap               & $d$      & $1.18\,\upmu$m             \\
		Young's modulus           & $E$      & $166\,$GPa                 \\
		Density                   & $\rho$   & $2332\,$kg\,m$^{-3}$       \\
		Nondimensional axial load & $N$      & $8.7$                      \\
		Dynamic viscosity of air  & $\mu$    & $1.983\times10^{-5}\,$Pa.s \\
		Structural damping        & $\gamma$ & $0.033$                    \\ \hline
	\end{tabular}
\end{table}

Unlike the microswitch of Section \ref{sec:validation}, this microbeam is wide compared with the gap, and its film is strongly coupled to the structure. At atmospheric pressure and zero bias, the air spring raises the first nondimensional natural frequency from $24.64$ to $326.1$, and twelve fluid modes lie below the first coupled mode, so that $\Lambda \ll 1$ (see Section \ref{sec:fluidmodes}). 

After a few simulations, it is found that between $5$ and $101.3$\,kPa, the quality factor of the first coupled mode remains close to $4$. At these pressures, the frequency response is nearly linear for small voltages, whereas large voltages produce a very wide pull-in band. The ambient pressure is therefore lowered to $p_a = 100$\,Pa. At this pressure and for $V_{DC}=2$\,V, the gas spring is around $12$\% of the stiffness of the first coupled mode with a quality factor $Q=13.1$. The slowest fluid mode is below it with $\Lambda \approx 0.5$. The film then relaxes on the timescale of the oscillation and carries its own dynamics. The Knudsen number $K_n \approx 56$ is far beyond the slip regime, and the effective viscosity \eqref{eq7} is used outside the range for which they proposed. The results of this section should therefore be read as a model-based study of the effect of the fluid modes on the dynamics, not as a quantitative prediction of the damping.

\subsection{Frequency response}\label{sec:frc}
\begin{figure}[H]
	\centering
	\includegraphics[width=0.6\columnwidth]{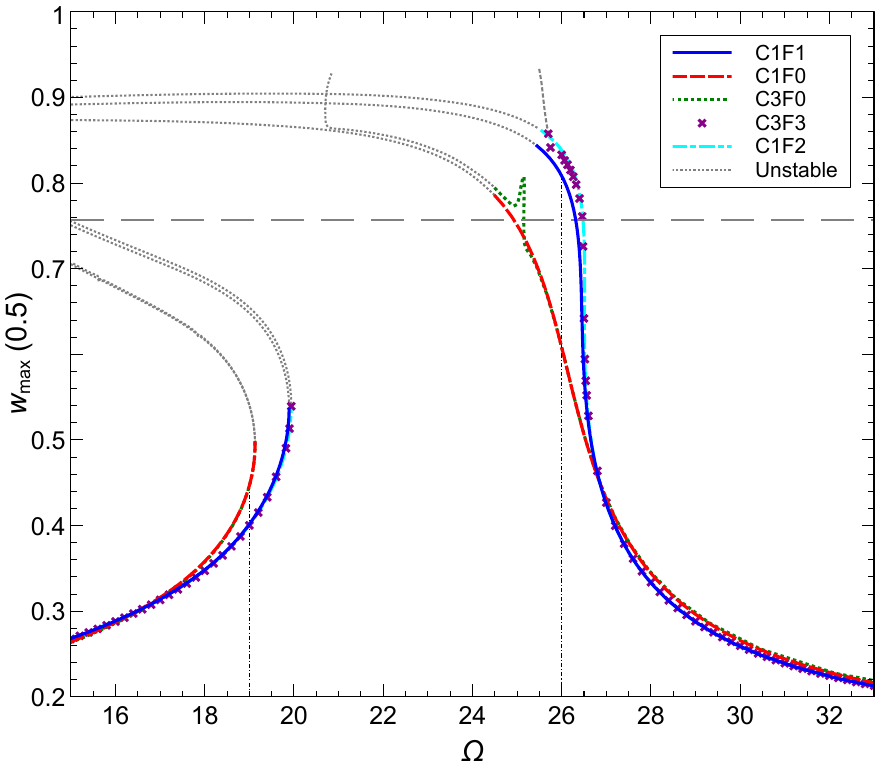}
	\caption{Frequency-response curves of the maximum midspan deflection for $V_{DC}=3.5$\,V, $v_{ac}=1$\,V and $p_a=100$\,Pa, computed with five reduced bases. Unstable orbits are shown in grey. The horizontal dashed line marks the unstable static deflection at $V_{RMS}$, and the vertical dash-dotted lines mark the frequencies $\Omega=19$ and $\Omega=26$ at which the basins of attraction are computed in Section \ref{sec:basins}.\label{fig:conv}}
\end{figure}

The microbeam is actuated with $V_{DC} = 3.5\,$V and $v_{ac} = 1\,$V, which corresponds to the softening case of \citet{najar2010nonlinear}, the static deflection and the reduced basis is computed at $V_{RMS} = 3.57\,$V. At these excitation voltages, the first coupled mode has a frequency of $23.81$ and a quality factor of $10.8$, and $5$ fluid modes lie below it. Figure \ref{fig:conv} shows the maximum steady state midspan deflection for five bases: C1F0, C3F0, C1F1, C1F2, and C3F3. The stability of the periodic solutions is determined from their Floquet multipliers, and the horizontal dashed line marks the unstable static deflection $u_{max} = 0.757$ at $V_{RMS}$.

The response consists of two branches separated by a frequency band in which no stable periodic orbit exists (pull-in band). The left branch is of the softening type and loses its stability at a cyclic-fold bifurcation, where a Floquet multiplier leaves the unit circle through $+1$. On the right branch, the orbits grow as the frequency decreases and cross the unstable static deflection without losing their stability. They lose it at a period-doubling bifurcation, where a Floquet multiplier leaves the unit circle through $-1$. The static saddle, shown for reference, therefore does not bound the stable orbits, and both edges of the band are set by bifurcations of the periodic solutions. In the classification of \citet{nayfeh2007dynamic}, these bifurcations correspond to the mechanism of dynamic pull-in that does not depend on the initial conditions.

The five bases are of two types: bases that include a fluid mode (C1F1, C1F2, and C3F3), coincide almost over the whole stable range; and bases without fluid mode (C1F0 and C3F0), coincide with each other but are different from the first type. The only exception occurs near the end of the right branch, where the C3F0 basis exhibits a narrow peak close to $\Omega = 25.2$ for C3F0 and to $\Omega = 25.7$ for C3F3. This peak results from a simultaneous resonance between the primary resonance of the first mode and the superharmonic resonance of the third mode of the microbeam, as identified by \citet{najar2010nonlinear} in the absence of squeeze film.

Adding coupled modes leaves the response unchanged, whereas a single fluid mode changes it and a second one adds nothing. In fact, the omission of the fluid mode displaces both edges of the band. The cyclic fold of the left branch moves from $\Omega=19.13$ without fluid mode to $19.89$, and the period-doubling bifurcation of the right branch from $\Omega=24.5$ to $25.4$. The band is thus shifted towards higher frequencies when the fluid mode is added. A model without fluid modes would therefore misplace the frequency window in which the microbeam can be switched by dynamic pull-in, at both ends. The Floquet multipliers characterize the stability of the orbits with respect to infinitesimal perturbations only. Their robustness with respect to perturbations of finite amplitude is examined in the next subsection.

\subsection{Basin of attraction of bounded motions}\label{sec:basins}
The local stability of the periodic orbits does not indicate whether a perturbation of finite amplitude, such as a mechanical shock or an electrical disturbance, drives the microbeam to pull-in. The global stability is assessed through the basin of attraction of bounded motions, that is, the set of initial conditions from which the response remains bounded.

For the C1F0 basis the reduced model has two real unknowns, and the basin is a domain of the plane $(w(0.5), \dot w(0.5))$. The C1F1 basis adds the real coordinate of the fluid mode, so that the phase space becomes three-dimensional and an initial condition in the plane $(w(0.5), \dot w(0.5))$ no longer defines the state of the system because the initial pressure must also be specified. The initial state is chosen here to represent a film at ambient pressure.

To simplify the notation, the coupled and fluid modes retained in the C1F1 basis are denoted by the subscripts $C$ and $F$. In the numbering of Equation \eqref{eq21}, the fluid mode is the first mode and the coupled mode the second. The modal coordinates are written $\eta_C = a_C + \mathrm{i}\, b_C$ and $\eta_F = a_F$, the right eigenvector of the coupled and fluid modes are ${\bf \Phi}_C$ and ${\bf \Phi}_F$,respectively. Evaluating the expansion \eqref{eq17} at $t=0$ at the midspan node, of index $m$, gives the initial deflection $w_m(0)$ and velocity $\dot{w}_m(0)$ of the midspan as
\begin{align}
	w_m(0) &= u_m + 2\left( a_C\, \Re(\Phi^{1}_{C,m}) - b_C\, \Im(\Phi^{1}_{C,m}) \right) + a_F\, \Phi^{1}_{F,m} \label{eqIC1} \\
	\dot{w}_m(0) &= 2\left( a_C\, \Re(\Phi^{3}_{C,m}) - b_C\, \Im(\Phi^{3}_{C,m}) \right) + a_F\, \Phi^{3}_{F,m} \label{eqIC2}
\end{align}
where $u_m$ is the static deflection of the midspan at $V_{RMS}$, the superscripts $1$ and $3$ denote the displacement and velocity partitions of the eigenvectors, and $w_m(0)$ is the total deflection. Equations \eqref{eqIC1} and \eqref{eqIC2} can be written as
\begin{equation} \label{eqIC3}
	2 \left[ \begin{array}{cc} \Re(\Phi^{1}_{C,m}) & -\Im(\Phi^{1}_{C,m}) \\ \Re(\Phi^{3}_{C,m}) & -\Im(\Phi^{3}_{C,m}) \end{array} \right] \left( \begin{array}{c} a_C \\ b_C \end{array} \right) =
	\left( \begin{array}{c} w_m(0) - u_m \\ \dot{w}_m(0) \end{array} \right) - a_F \left( \begin{array}{c} \Phi^{1}_{F,m} \\ \Phi^{3}_{F,m} \end{array} \right)
\end{equation}
Since the velocity partition of an eigenvector satisfies ${\bf \Phi}^{\bf{3}}_C = \lambda_C {\bf \Phi}^{\bf{1}}_C$, with $\lambda_C = \varsigma_C + \mathrm{i}\,\omega_{d,C}$, the determinant of the left-hand side matrix in Equation \eqref{eqIC3} is $-4\,\omega_{d,C} \left| \Phi^{1}_{C,m} \right|^2$. The coordinates of the coupled mode are given by
\begin{equation} \label{eqIC4}
	\left( \begin{array}{c} a_C \\ b_C \end{array} \right) = {\bf p}_0 + a_F\, {\bf p}_1, 
\end{equation}
where
$${\bf p}_0=\frac12 \left[ \begin{array}{cc} \Re(\Phi^{1}_{C,m}) & -\Im(\Phi^{1}_{C,m}) \\ \Re(\Phi^{3}_{C,m}) & -\Im(\Phi^{3}_{C,m}) \end{array} \right]^{-1} \left( \begin{array}{c} w_m(0) - u_m \\ \dot{w}_m(0) \end{array} \right) $$ 
and 
$${\bf p}_1 = - \frac12 \left[ \begin{array}{cc} \Re(\Phi^{1}_{C,m}) & -\Im(\Phi^{1}_{C,m}) \\ \Re(\Phi^{3}_{C,m}) & -\Im(\Phi^{3}_{C,m}) \end{array} \right]^{-1} \left( \begin{array}{c} \Phi^{1}_{F,m} \\ \Phi^{3}_{F,m} \end{array} \right) $$
For the C1F0 basis, which contains no fluid mode, $a_F = 0$, and Equation \eqref{eqIC4} reduces to $(a_C, b_C) = {\bf p}_0$, which defines the initial state completely.

For the C1F1 basis, the coordinate $a_F$ remains to be determined. Substituting Equation \eqref{eqIC4} into the pressure partition of the expansion \eqref{eq17}, the initial pressure is an affine function of $a_F$,
\begin{equation} \label{eqIC5}
	{\bf z}^{\bf 2}(0) = {\bf P}_0 + a_F\, {\bf P}_1,
\end{equation}
where ${\bf P}_0 = 2\left( p_{0,1}\, \Re\left( {\bf \Phi}^{\bf{2}}_C\right) - p_{0,2}\, \Im\left( {\bf \Phi}^{\bf{2}}_C\right) \right)$, ${\bf P}_1 = 2\left( p_{1,1}\, \Re\left( {\bf \Phi}^{\bf{2}}_C\right) - p_{1,2}\, \Im\left( {\bf \Phi}^{\bf{2}}_C\right) \right) + {\bf \Phi}^{2}_{F}$, $p_{0,k}$ and $p_{1,k}$ are the components of ${\bf p}_0$ and ${\bf p}_1$. For an ambient pressure condition ${\bf z}^{\bf 2}(0) = {\bf 0}$, which cannot be satisfied for arbitrary $(w_m(0), \dot{w}_m(0))$ with a single free coordinate. The coordinate $a_F$ is therefore chosen to minimize the norm of the initial pressure, which gives
\begin{equation} \label{eqIC6}
	a_F = -\frac{{\bf P}_0 \cdot {\bf P}_1}{{\bf P}_1 \cdot {\bf P}_1}
\end{equation}
the coordinates $a_C$ and $b_C$ then following from Equation \eqref{eqIC4}. This choice defines the film closest to ambient pressure compatible with the prescribed motion of the midspan. All the initial conditions are imposed at $t=0$, i.e.\ at the same phase of the excitation, for both bases.

The phase plane is subdivided into a grid of $400\times400$ points, used as an initial condition for the integration of the reduced model over 80 periods of excitation. An initial condition is counted as leading to pull-in when $w(0.5)$ exceeds $0.99$. The basins are computed with the C1F0 and C1F1 bases at the same voltage, for two excitation frequencies located around the pull-in band of Figure \ref{fig:conv}, $\Omega = 19$ and $\Omega = 26$.

\begin{table}[H]
	\centering
	\caption{Integrity measures of the basins of attraction of bounded motions for $V_{DC}=3.5$\,V and $p_a=100$\,Pa. In all cases the basin contains a single attractor. $w_{max}$: maximum midspan deflection of the attractor; $\omega_1$: first coupled frequency at the operating point; IF: integrity factor divided by IF$_0$, radius of the largest circle contained in the safe basin; LIM: local integrity measure divided by LIM$_0$, radius of the largest circle centred at the attractor and contained in the safe basin; GIM: global integrity measure divided by GIM$_0$, area of the safe basin. The measure without excitation for the same basis are: IF$_0=0.570$, LIM$_0=0.526$ and GIM$_0=1.242$ for C1F0, and IF$_0=0.657$, LIM$_0=0.541$ and GIM$_0=1.733$ for C1F1. All measures are evaluated in the normalized phase plane $(w(0.5),\, \dot w(0.5)/\omega_1)$ on a grid of $400\times400$ initial conditions.}
	\label{tab:basins}
	\begin{tabular}{cccccccc}
		\hline
		$\Omega$ & $v_{ac}$ (V) & Basis & $w_{max}$ & $\omega_1$ & IF & LIM & GIM \\ \hline
		19    &      1       & C1F0  &   0.447   &   23.810   &   0.18    &    0.09     &    0.08     \\
		19    &      1       & C1F1  &   0.402   &   23.810   &   0.39    &    0.20     &    0.24     \\ \hline
		26    &      1       & C1F0  &   0.608   &   23.810   &   0.96    &    0.30     &    0.91     \\
		26    &      1       & C1F1  &   0.809   &   23.810   &   1.10    &    0.10     &    1.12     \\
		26    &     1.05     & C1F0  &   0.638   &   23.797   &   0.89    &    0.26     &    0.85     \\
		26    &     1.05     & C1F1  &   0.828   &   23.797   &   0.85    &    0.08     &    0.94     \\
		26    &     1.1      & C1F0  &   0.666   &   23.782   &   0.83    &    0.22     &    0.78     \\
		26    &     1.1      & C1F1  &   0.843   &   23.782   &   0.12    &    0.06     &    0.34     \\ \hline
	\end{tabular}
\end{table}

The robustness of the response to finite perturbations is quantified by three measures, summarized in Table \ref{tab:basins} \citep{Soliman_Thompson_1989,Lenci_Rega_2003,ruzziconi2013electrically}. The integrity factor (IF) is the radius of the largest circle contained in the safe basin and measures the compact core of the basin. The local integrity measure (LIM) is the radius of the largest circle centred at the attractor, and measures the smallest perturbation of the operating orbit that leads to pull-in. The global integrity measure (GIM) is the area of the safe basin. All the basins are computed for initial conditions imposed at $t=0$, i.e.\ at the same phase of the excitation, the attractor is represented by its position at that phase. Each measure is normalized by its value without excitation for the same basis. The attractor is the static equilibrium, whose basin is bounded by the stable manifold of the saddle. Its size depends on the choice of the basis. 

Table \ref{tab:basins} shows that by including the fluid mode the IF increases from $0.570$ to $0.657$, the LIM from $0.526$ to $0.541$, and the GIM from $1.242$ to $1.733$. With the fluid mode, the pressure is no longer only driven by the motion of the microbeam, and the gas compressed by a rapid approach of the microbeam can relax only on the timescale of the fluid mode, comparable here to the period of oscillation. The resulting transient pressure opposes the the microbeam motion near the substrate, so that larger initial perturbations are needed to cross the saddle.

\begin{figure}[H]
	\centering
	\subfloat[][C1F0: Attractor $w_{max}=0.608$, $t_{max}=0.878$.\label{figBass26a}]
	{\scalebox{0.95}{\includegraphics[width=0.8\textwidth]{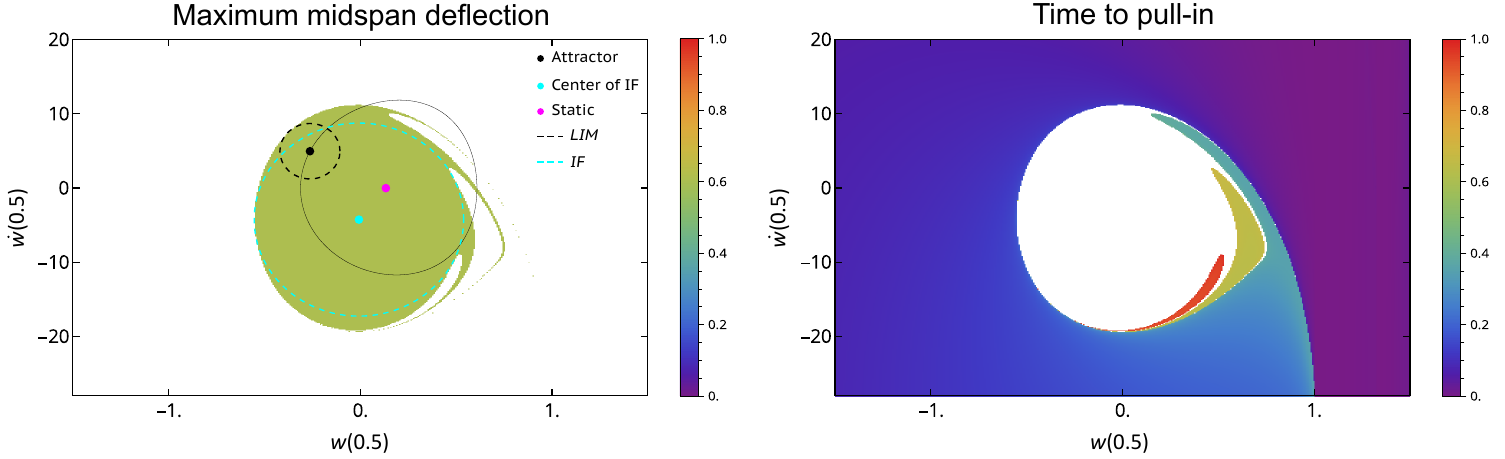}}}\\
	\subfloat[][C1F1: Attractor $w_{max}=0.809$, $t_{max}=1.431$.\label{figBass26b}]
	{\scalebox{0.95}{\includegraphics[width=0.8\textwidth]{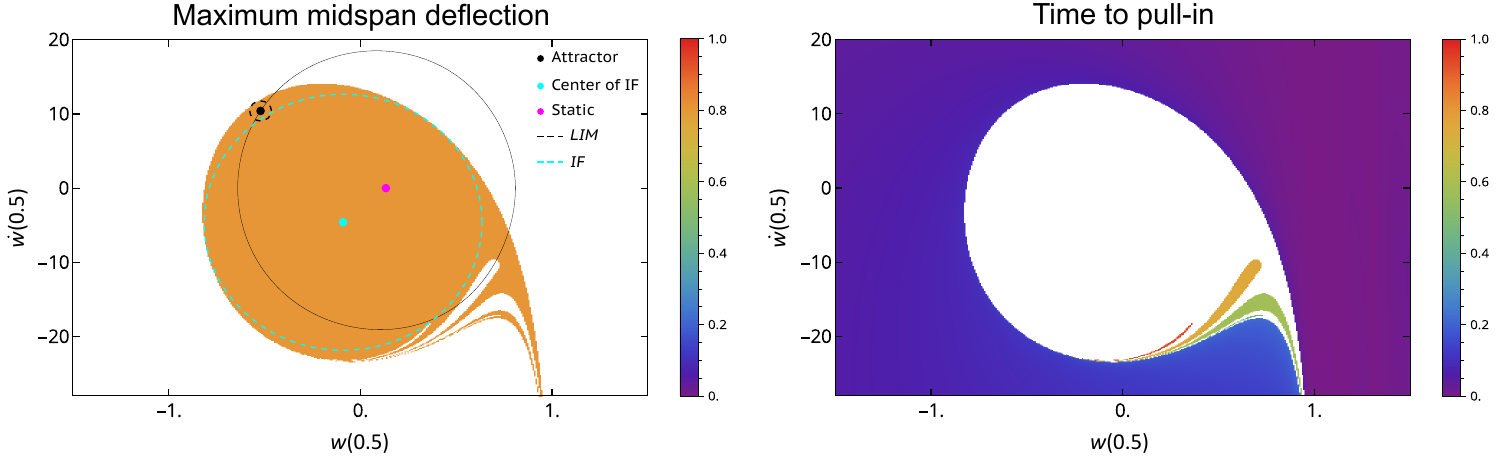}}}
	\caption{Basins of attraction of bounded motions at $\Omega=26$ for $V_{DC}=3.5$\,V, $v_{ac}=1$\,V and $p_a=100$\,Pa. Left: maximum midspan deflection of the attractor reached from each initial condition, right: time to pull-in normalized by its maximum value $t_{max}$. White regions correspond to pull-in on the left and to bounded motions on the right. Black point and dashed black circle: attractor at phase 0 and LIM, cyan point and dashed cyan circle: centre of the IF circle and IF, magenta point: static deflection at $V_{RMS}$. The thin grey curve is the attracting orbit, only its point at phase 0 (black) is represented in the basin, which describes initial conditions imposed at that phase.}
	\label{figBass26}
\end{figure}

Figure \ref{figBass26} compares the basins obtained at $\Omega=26$ with and without the fluid mode. Each basin contains a single attractor with all bounded initial conditions converging to the same periodic orbit. However, as shown by the corresponding frequency response curve, the two bases do not predict the same orbit. Without the fluid mode, the orbit has a maximum deflection $0.608$, whereas with it the orbit is at $0.809$ of maximum deflection. It is also deduced that for a base with fluid mode, the safe basin is larger and its integrity factor and area exceeding even those of the unforced equilibrium with IF$=1.10$ and GIM$= 1.12$, against $0.96$ and $0.91$ without the fluid mode. Also in this base, the attractor is situated close to the basin boundary so that its local measure drops to LIM$ = 0.10$, against $0.30$ without the fluid mode. The larger orbit predicted with the fluid mode is thus surrounded by a wide safe region, but a perturbation three times smaller than in the model without fluid mode is enough to take it to pull-in.

\begin{figure}[H]
	\centering
	\subfloat[][C1F0: Attractor $w_{max}=0.447$, $t_{max}=7.629$.\label{figBass19a}]
	{\scalebox{0.95}{\includegraphics[width=0.8\textwidth]{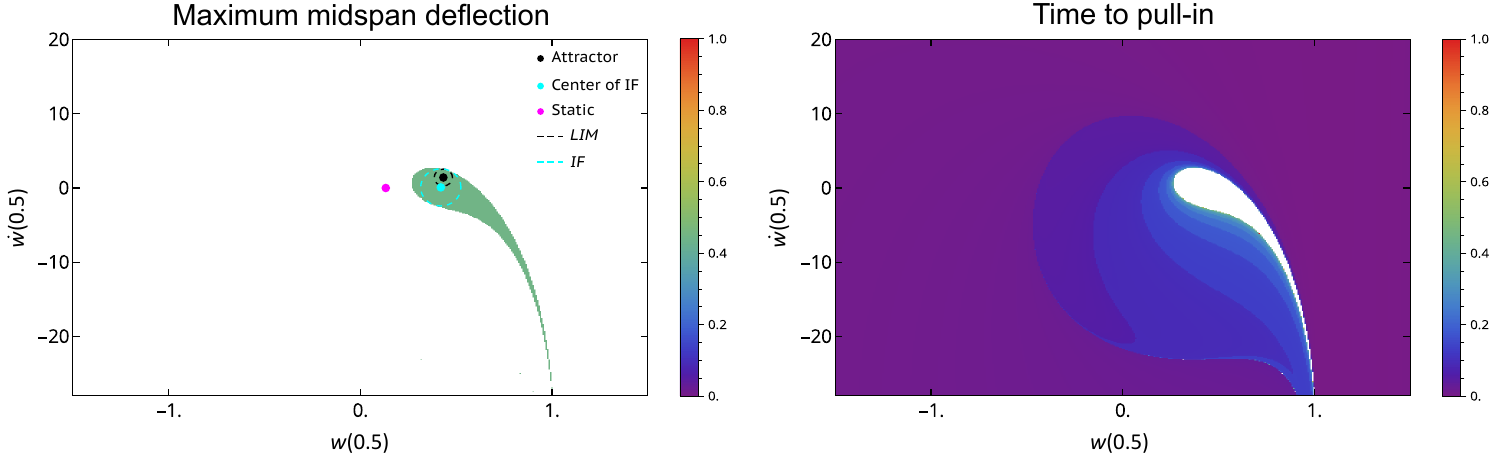}}}\\
	\subfloat[][C1F1: Attractor $w_{max}=0.402$, $t_{max}=3.005$.\label{figBass19b}]
	{\scalebox{0.95}{\includegraphics[width=0.8\textwidth]{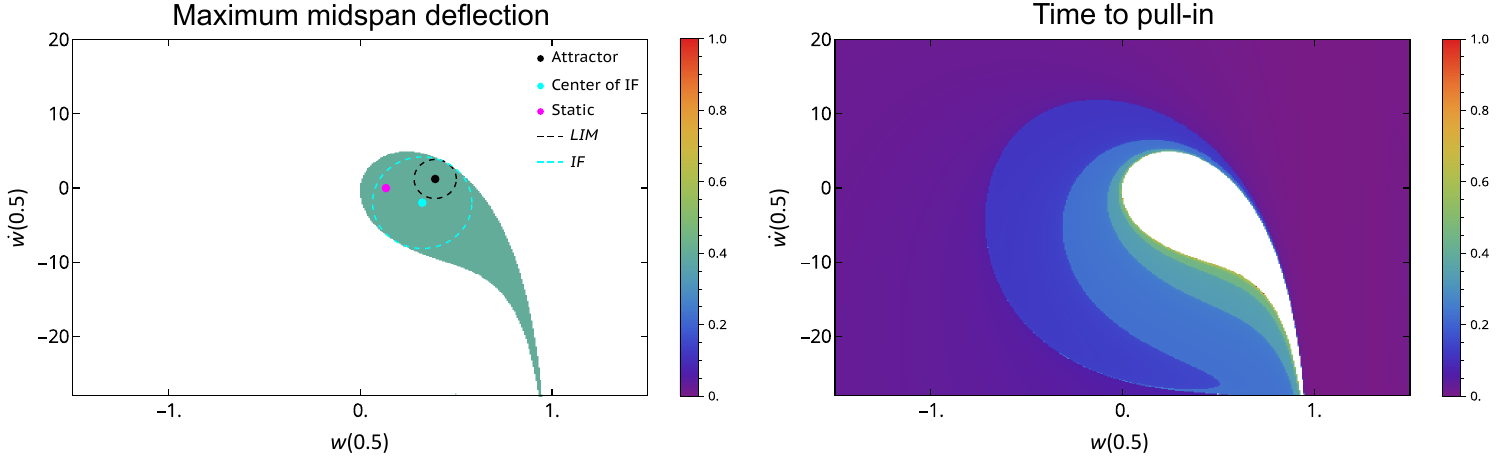}}}
	\caption{Basins of attraction of bounded motions at $\Omega=19$ for $V_{DC}=3.5$\,V, $v_{ac}=1$\,V and $p_a=100$\,Pa. Same representation as in Figure \ref{figBass26}.}
	\label{figBass19}
\end{figure}

At $\Omega=19$, close to the cyclic fold of the left branch, the basins are small and elongated towards the saddle, as depicted by Figure \ref{figBass19}, the attracting orbits is situated on the lower branch with maximum deflections of $0.447$ and $0.402$. Here the three measures have the same behavior: without the fluid mode the normalized IF is reduced to $0.18$ of its unforced value, the LIM to $0.09$ and the GIM to $0.08$, against $0.39$, $0.20$ and $0.24$ with the fluid mode. Since the cyclic-fold occurs at $\Omega \approx 19.13$ for C1F0 and $19.89$ for C1F1, the excitation frequency is therefore much closer to the bifurcation point in the case C1F0. As a result, the stable orbit approaches the unstable one and the domain of bounded initial conditions contracts.

\begin{figure}[H]
	\centering
	\subfloat[][$v_{ac}=1.05$\,V: Attractor $w_{max}=0.828$, $t_{max}=3.268$.\label{figBassACa}]
	{\scalebox{0.95}{\includegraphics[width=0.8\textwidth]{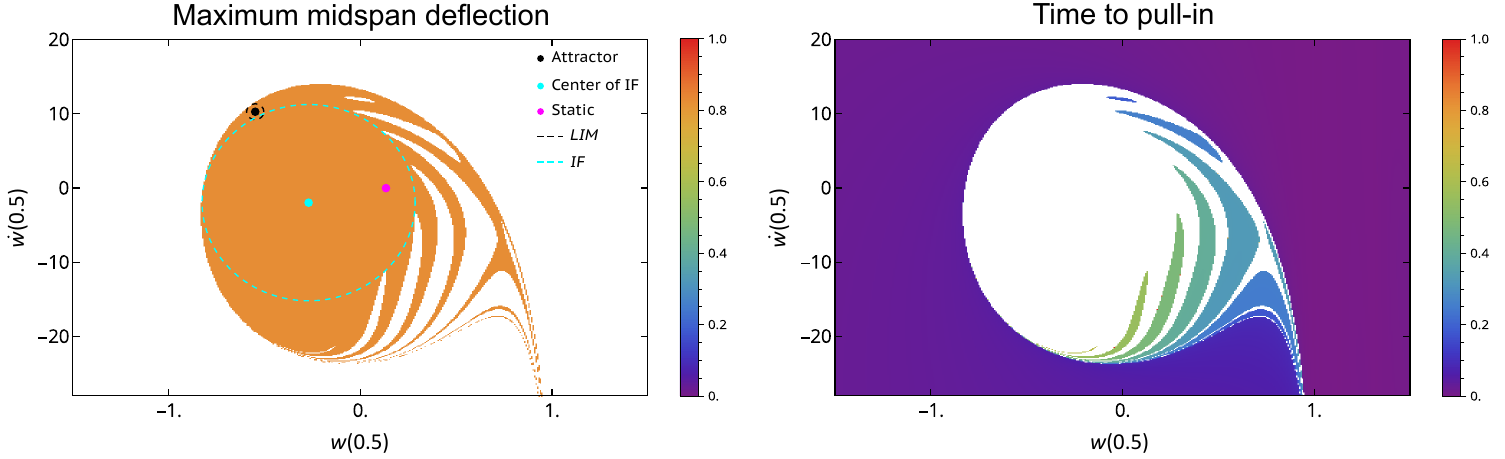}}}\\
	\subfloat[][$v_{ac}=1.1$\,V: Attractor $w_{max}=0.843$, $t_{max}=10.749$.\label{figBassACb}]
	{\scalebox{0.95}{\includegraphics[width=0.8\textwidth]{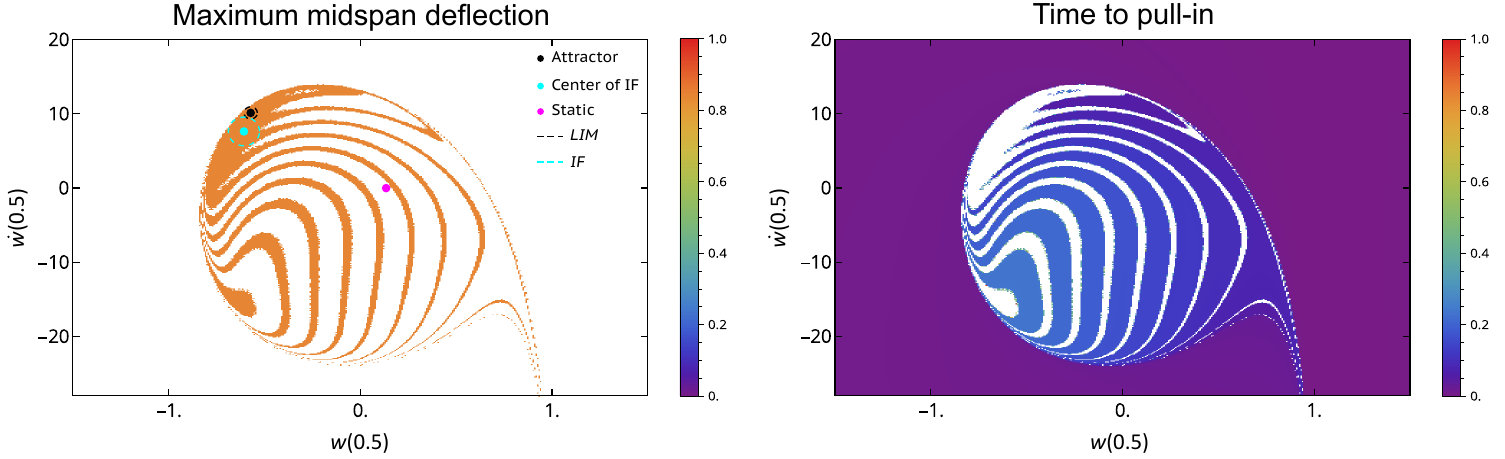}}}
	\caption{Erosion of the basin of attraction with the excitation amplitude at $\Omega=26$, $V_{DC}=3.5$\,V and $p_a=100$\,Pa, computed with the C1F1 basis. The case $v_{ac}=1$\,V is shown in Figure \ref{figBass26b}. Same representation as in Figure \ref{figBass26}.}
	\label{figBassAC}
\end{figure}

The effect of the excitation amplitude is illustrated in Figure \ref{figBassAC} for the C1F1 basis at $\Omega=26$. As $v_{ac}$ increases from $1$ to $1.1$\,V, the attractor maximum deflection increases from $0.809$ to $0.843$, but the basin is invaded by fractal tongues of initial conditions leading to pull-in. IF falls from $1.10$ to $0.85$ at $v_{ac}=1.05$\,V and to $0.12$ at $v_{ac}=1.1$\,V, and GIM from $1.12$ to $0.94$ and then to $0.34$ (Table \ref{tab:basins}). At $v_{ac}=1.1$\,V the compact part of the basin has almost disappeared, and the safe initial conditions are confined to thin interleaved bands, so that the outcome becomes extremely sensitive to the initial state. The LIM decreases much less, from $0.10$ to $0.08$ and $0.06$ because the attractor was already close to the boundary at $v_{ac}=1$\,V. The longest pull-in times increase from $1.431$ to $3.268$ and then to $10.749$, that is from about $6$ to $44$ periods of excitation, and are reached along the interleaved bands. An increase of $10$\% of the excitation destroys the stability of an orbit that remains locally stable, which is the mechanism of erosion described by \citet{najar2010dynamics} in the absence of film, and defines a practical pull-in threshold well below the local stability limit of the orbit. It also corresponds to the third mechanism of dynamic pull-in identified by \citet{nayfeh2007dynamic}, in which small transient deviations from a stable periodic orbit lead to pull-in.

The model without fluid mode does not capture this collapse. Over the same range of amplitudes its basin erodes slowly and regularly, the IF decreasing from $0.96$ to $0.89$ and $0.83$, the LIM from $0.30$ to $0.26$ and $0.22$, and the GIM from $0.91$ to $0.85$ and $0.78$. The description of the two models by integrity factor is reversed between $v_{ac}=1$ and $1.05$\,V, and at $v_{ac}=1.1$\,V the model without fluid mode predicts an integrity factor seven times larger than that obtained with the fluid mode.

The fluid mode therefore affects the global stability in a way that depends on the operating point and on the measure considered. Close to the cyclic-fold, it increases the robustness of the response according to all three measures. Away from the fold, it predicts a larger safe basin at moderate excitation, but an attractor located close to its boundary, and an abrupt erosion of the basin as the excitation increases. A model without fluid mode would overestimate the robustness of the operating orbit at all the amplitudes considered at $\Omega=26$, and miss the collapse of the basin that defines the practical pull-in threshold.

\subsection{Dynamic pull-in maps}\label{sec:maps}
The frequency-response curves and the basins of attraction describe the response for a given excitation. For a microswitch actuated by dynamic pull-in, the relevant question is to know which excitations drive the microbeam to pull-in from its state at rest. Following the approach used for microswitches by \citet{sassi2018square}, and used by \citet{Gusso_Ujevic_2025} to compare damping models, this question is answered by maps in the plane of the excitation parameters $(\Omega, v_{ac})$ at $V_{DC}=3.5$\,V and $p_a=100$\,Pa. Each point of the map is obtained by integrating the reduced model over $80$ periods of excitation from the same initial state, the microbeam is at rest at its static deflection and the film at ambient pressure, all the modal coordinates are zero at $t=0$. As in Section \ref{sec:basins}, pull-in is declared when the midspan deflection exceeds $0.99$. The plane is covered by $400 \times 300$ points, for $0.01 \le \Omega \le 32$ and $0 \le v_{ac} \le 1.5$\,V.

\begin{figure}[H]
	\centering
	\subfloat[][C1F1, $w_{max}=0.890$, $t_{max}=34.191$.\label{figMapC1F1}]
	{\includegraphics[width=0.95\textwidth]{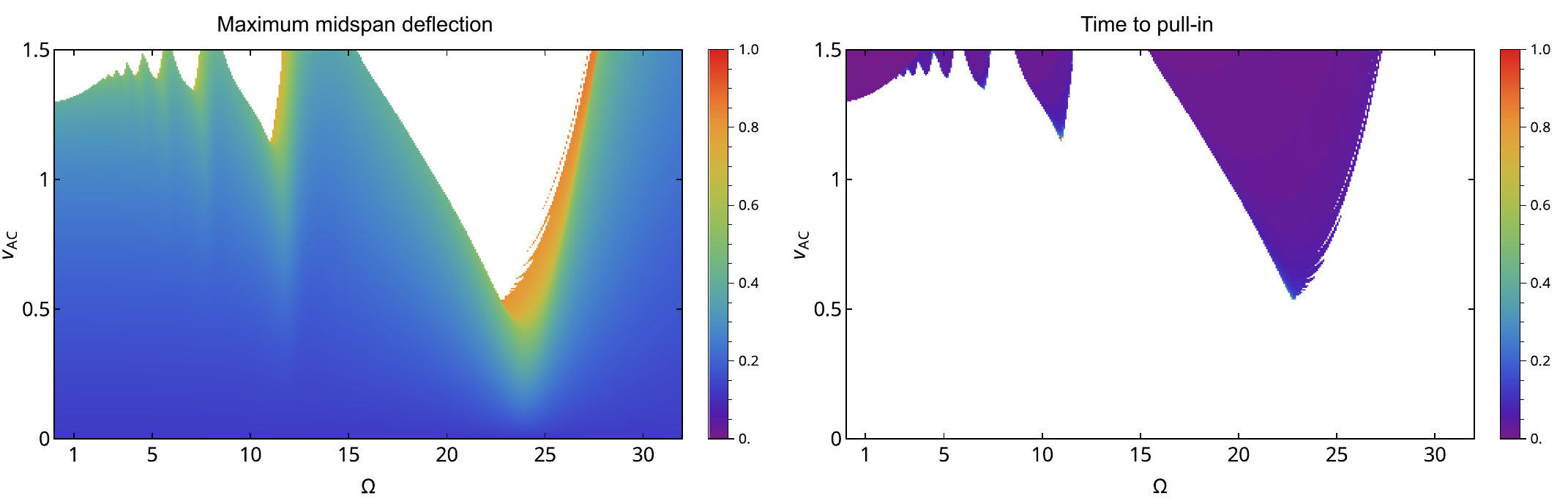}}\\
	\subfloat[][C1F0, $w_{max}=0.840$, $t_{max}=18.951$.\label{figMapC1F0}]
	{\includegraphics[width=0.95\textwidth]{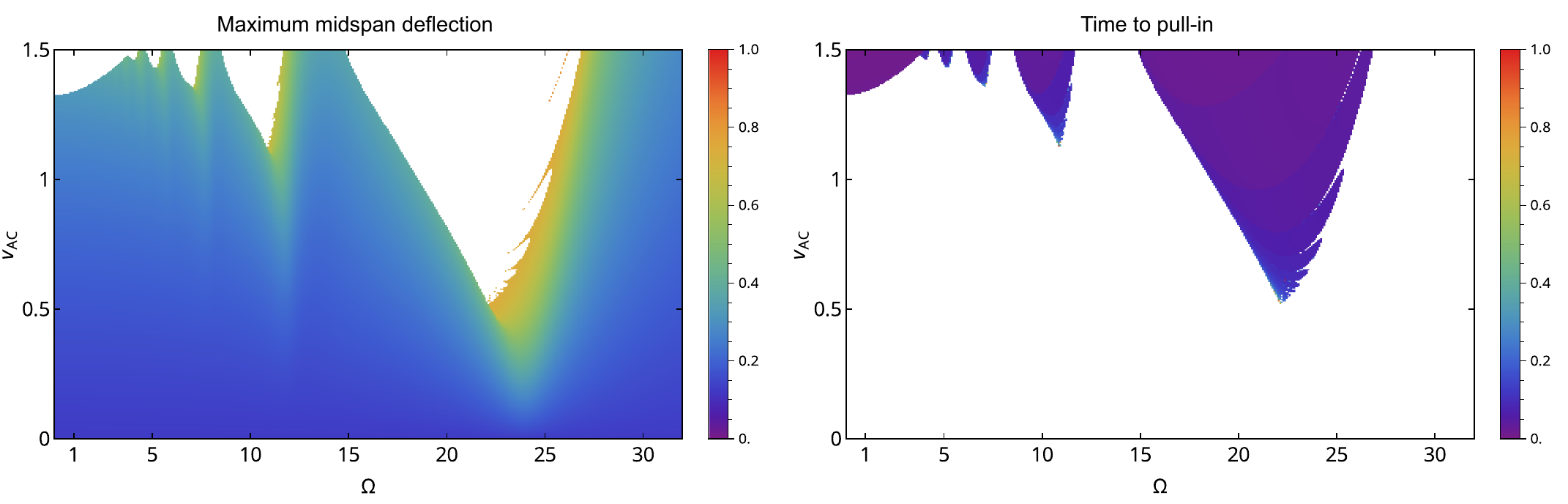}}
	\caption{Dynamic pull-in maps in the plane $(\Omega, v_{ac})$ for $V_{DC}=3.5$\,V and $p_a=100$\,Pa, the microbeam starting at rest from its static deflection. Left, maximum normalized midspan deflection of the bounded responses. Right, time to pull-in normalized by its maximum value. White regions correspond to pull-in on the left and to bounded responses on the right.}
	\label{figMaps}
\end{figure}

The two maps of Figure \ref{figMaps} share several features. At low frequency the voltage varies slowly enough for the microbeam to follow it, and pull-in occurs when the peak voltage $V_{DC}+v_{ac}$ reaches the static pull-in voltage, which gives a threshold close to $v_{ac} = 1.3$\,V for both bases. This limit does not depend on the film, which plays no role in the static problem, and its coincidence for the two bases confirms the consistency of the computations. Between $\Omega \approx 2$ and $4$ the threshold rises slightly above this value, the voltage exceeding the static limit during a short fraction of the period for the microbeam to escape. At higher frequencies, a sequence of narrow switching windows opens. These frequency bands are close to integer fractions of the natural frequency of the fundamental mode, corresponding to the superharmonic resonance of order two near $\Omega \approx 11$. The threshold of these windows increases with their order, in agreement with the observations of \citet{sassi2018square} in the absence of film. The primary switching region has the shape of a V, whose tip defines the minimum voltage for switching, $v_{ac}=0.525$\,V at $\Omega=22.08$ without the fluid mode and $v_{ac}=0.535$\,V at $\Omega=22.83$ with it.

The fluid mode shifts the whole primary region towards higher frequencies, while leaving the minimum switching voltage nearly unchanged. At $v_{ac}=1$\,V the switching region extends from $\Omega=18.56$ to $25.20$ without the fluid mode, and from $\Omega=19.43$ to $25.74$ with it. The frequency response of Figure \ref{fig:conv} showed this displacement of the edges of the pull-in band at a single amplitude, $v_{ac}=1$\,V. The maps show that it holds over the whole range of amplitudes considered.

For both models, the left edge of the switching region lies below the corresponding cyclic fold, at $\Omega = 18.56$ instead of $19.13$ without the fluid mode, and at $\Omega = 19.43$ instead of $19.89$ with it. In these frequency bands a stable periodic orbit exists, yet the microbeam started from rest does not reach it and pulls in during the transient. This transient-driven dynamic pull-in, described by \citet{nayfeh2007dynamic}, cannot be captured by the frequency-response curves, which only follow the periodic orbits.

In Figure \ref{figMaps}, the right edge of the primary region is irregular for both bases, thin tongues of bounded responses penetrating the switching region, so that near this edge the outcome alternates between switching and bounded response under small changes of the frequency or of the amplitude. The fluid mode changes the shape of these tongues rather than their existence. Without it, the tongues are shorter and wider, whereas with it, they become longer and thinner. Near the right edge of the region, the outcome of the actuation is therefore sensitive to small variations of the excitation in both bases, and the fluid mode modifies the geometry of the zone in which this sensitivity occurs.

%%%%%%%%%%%%%%%%%%%%%%%%%%%%%%%%%%%%%%%%%%%%%%%%%%%%%%%%%%%%%%%%%%%%%%%%%%%%%%%%%%%%%%%%%%%%%%%%%%%%%%%%%%%%%%%%%%%%%%%%%%%%%
\section{Conclusion}\label{sec:conclusions}
A reduced-order model based on the complex modes of the coupled microbeam--air film problem has been extended to the nonlinear periodic response, the local and global stability, and the dynamic pull-in of electrostatically actuated clamped--clamped microbeams. The model was validated against the microswitch of \citet{sassi2018square}. It reproduces the measured static pull-in voltage and a basis with only one coupled mode and one fluid relaxation mode, reproduces the frequency response of the fully coupled reference model. Bases without fluid mode, in contrast, overestimate the upper branch and misrepresent its amplitudes near resonance, whatever the number of coupled modes they retain. The remaining discrepancy with the experiment was shown to originate in the film model and not in the modal truncation.

The fluid relaxation modes were characterized in closed form with a uniform gap assumption. In this case, their eigenvalues follow from the linearized Reynolds equation alone, and the ratio of the slowest one to the first natural frequency defines a parameter that distinguishes quasi-static films from films that carry their own dynamics.

For the clamped--clamped microbeam of \citet{tilmans1994electrostatically} at $p_a=100$\,Pa, where the slowest fluid mode relaxes on the timescale of the oscillation, a single fluid mode was found to be necessary and sufficient. In the frequency response, the fluid mode displaces both branches of the pull-in band. The cyclic-fold of the left branch moves from $\Omega=19.13$ to $19.89$, and the period-doubling bifurcation of the right branch from $24.5$ to $25.4$.

The fluid mode also modifies the global stability. It enlarges the basin of the unforced equilibrium and, near the cyclic-fold, increases all integrity measures by a factor of two to three. Away from it, a larger safe basin is obtained, however the local integrity measure is three to four times smaller than without the fluid mode. As the excitation increases by $10$\%, this basin erodes abruptly, its normalized integrity factor falling from $1.10$ to $0.12$, whereas the basin obtained without the fluid mode erodes slowly and regularly and misses this collapse.

In the plane of the excitation parameters, the fluid mode shifts the whole region of dynamic pull-in towards higher frequencies, while leaving the minimum switching voltage nearly unchanged, at about $0.53$\,V for $V_{DC}=3.5$\,V. For both models, the left edge of this region is located below the cyclic fold, the microbeam started from rest pulling in during the transient although a stable orbit exists. Near the right edge, thin tongues of bounded responses penetrate the switching region in both models, the fluid mode making them longer and thinner.

A model without fluid mode would therefore misplace the frequency window in which the microbeam can be switched, overestimate the robustness of its operating orbit, and miss the practical pull-in threshold set by the erosion of the basin. These errors are of direct consequence for the design of microswitches actuated by dynamic pull-in.

These results were obtained at an ambient pressure for which the Knudsen number lies far beyond the slip regime, and they should be read as a model-based study of the role of the fluid modes rather than as a quantitative prediction of the damping. The basins were computed at a single phase of the excitation, and the dynamic pull-in maps from a single initial state. Further work will address the extension of the analysis to clamped--free microbeams, an a priori criterion for selecting the fluid modes, and experimental validation at reduced pressure, where the film carries its own dynamics.

%%%%%%%%%%%%%%%%%%%%%%%%%%%%%%%%%%%%%%%%%%%%%%%%%%%%%%%%%%%%%%%%%%%%%%%%%%%%%%%%%%%%%%%%%%%%%%%%%%%%%%%%%%%%%%%%%%%%%%%%%%%%%
\appendix

\section{DQM and LaDQM matrices} \label{ApI}

\begin{align}
	& A_{ij}^{(1)} =  {\frac{{\prod\limits_{v = {0};v \ne i}^{{n+1}} {({\xi _i} - {\xi _v})} }}{{({\xi _i} - {\xi _j})\prod\limits_{v = {0};v \ne j}^{{n+1}} {({\xi _j} - {\xi _v})} }}}  && {0} \le j \le {n+1}, \quad j \ne i \label{Ap01} \\
	& A_{ij}^{(k)} = k\left( A_{ii}^{(k-1)}A_{ij}^{(1)} - \frac{A_{ij}^{(k-1)}}{\xi_i - \xi_j} \right) && 0 \le j \le n+1, \quad j \ne i \\
	& A_{ii}^{(k)} = -\sum\limits_{v=0;\,v\ne i}^{n+1} A_{iv}^{(k)} && 0 \le i \le n+1 \\
	&	&& k = 2,3, \ldots \nonumber
\end{align}

\begin{align}
	& B_{ij}^{(1)} =  {\frac{{\prod\limits_{v = {1};v \ne i}^{{n}} {({\xi _i} - {\xi _v})} }}{{({\xi _i} - {\xi _j})\prod\limits_{v = {1};v \ne j}^{{n}} {({\xi _j} - {\xi _v})} }}}  && {1} \le j \le {n}, \quad j \ne i \label{Ap02} \\
	& B_{ij}^{(k)} = k\left( B_{ii}^{(k-1)}B_{ij}^{(1)} - \frac{B_{ij}^{(k-1)}}{\xi_i - \xi_j} \right) && 1 \le j \le n, \quad j \ne i \\
	& B_{ii}^{(k)} = -\sum\limits_{v=1;\,v\ne i}^{n} B_{iv}^{(k)} && 1 \le i \le n \\
	&	&& k = 2,3, \ldots \nonumber
\end{align}

\begin{equation}\label{eqNC}
	T_i=\int_{0}^{1} \prod_{j=1;j\neq i}^{n}\frac{x-\xi_j}{\xi_i-\xi_j}dx
\end{equation}

\section{Transfer and condensed derivative operators} \label{ApII}

The pressure values retained in ${\bf z^2}$ are ordered with the node $x_i$, $i = 2, \dots, n-1$, in the outer loop and the transverse nodes $j = 2, \dots, \frac{n+1}{2}$ in the inner loop, the same ordering is used for the complete $n \times n$ grid of the film. With $n_y = (n-1)/2$ retained nodes $y_j$ for each $x_i$, the operator
\begin{equation}
	{\bf E} = {\bf I}_{n_w} \otimes {\bf 1}_{n_y},
\end{equation}
of dimension $n_p \times n_w$, maps a quantity defined at the nodes $x_i$ of the microbeam onto the nodes $(x_i, y_j)$ of the film. Every node $(x_i, y_j)$ receives the value of that quantity at $x_i$, for whatever $y_j$ is. This reflects the fact that the deflection, and hence the gap, is uniform across the width. Here $\otimes$ denotes the Kronecker product. Conversely, the operator
\begin{equation}
	{\bf G} = {\bf I}_{n_w} \otimes {\bf g}^T,
	\qquad
	g_j =
	\begin{cases}
		T_j + T_{n+1-j}, & j = 2, \dots, \dfrac{n-1}{2},\\[4pt]
		T_{\frac{n+1}{2}}, & j = \dfrac{n+1}{2},
	\end{cases}
\end{equation}
of dimension $n_w \times n_p$, evaluates the transverse integral at across the width at each $x_i$, $\left({\bf G}\,{\bf z^2}\right)_i = \sum_{j} g_j\, p_{ij} \approx \int_0^1 p(x_i,y)\,dy$, where $T_j$ are the quadrature weights of Equation \eqref{eqNC}. Since only half of the film is retained, the value at each omitted node $n+1-j$ is replaced by that of its symmetric counterpart $j$. The center-line node is its own counterpart and keeps its weight, and the edge nodes, where the pressure vanishes, do not contribute.

The differential quadrature sums of \ref{ApI} extend over all the nodes of the grid, including the boundary and fictitious nodes eliminated through the boundary conditions and, in the $y$-direction, the nodes omitted by symmetry. Since all these conditions are linear, the complete nodal fields are linear functions of the retained values. Let ${\bf R}_p$ ($n^2 \times n_p$) be the operator that reconstructs the pressure at all nodes of the film from the retained values, and ${\bf S}_p$ ($n_p \times n^2$) the operator that selects the retained nodes, let ${\bf R}_w$ ($(n+2) \times n_w$) and ${\bf S}_w$ ($n_w \times (n+2)$) be the corresponding operators for the microbeam. The condensed derivative matrices are then
\begin{equation}
	{\bf a}^{(k)} = {\bf S}_w {\bf A}^{(k)} {\bf R}_w,
	\qquad
	{\bf b}_x^{(k)} = {\bf S}_p\left({\bf B}^{(k)} \otimes {\bf I}_n\right){\bf R}_p,
	\qquad
	{\bf b}_y^{(k)} = {\bf S}_p\left({\bf I}_n \otimes {\bf B}^{(k)}\right){\bf R}_p,
\end{equation}
square of dimension $n_w$ and $n_p$ respectively. Since both directions of the film are discretized on the same grid with the same number of points, ${\bf b}_x^{(k)}$ and ${\bf b}_y^{(k)}$ are built from the same weighting coefficients and differ only in the direction of summation and in the boundary conditions incorporated into ${\bf R}_p$.

\end{document}